\documentclass[preprint,12pt]{elsarticle}

\usepackage{amsmath}
\usepackage{amssymb}
\usepackage{bm}
\usepackage{amsfonts}
\usepackage{setspace}
\usepackage{optidef}
\usepackage{tabularx}
\usepackage{amsthm}
\usepackage{comment}
\usepackage{mathrsfs}
\usepackage{mathtools,xparse}

\usepackage[pdfpagemode={UseOutlines},bookmarks=true,bookmarksopen=true,
   bookmarksopenlevel=0,bookmarksnumbered=true,hypertexnames=false,
   colorlinks,linkcolor={blue},citecolor={blue},urlcolor={red},
   pdfstartview={FitV},unicode,breaklinks=true]{hyperref}

\usepackage[nameinlink]{cleveref}

\newcommand{\supplementary }{%
    \hyperref[sec:supplementary]{Supplementary Materials}%
}

\usepackage{booktabs} 

\usepackage{multirow}
\usepackage{makecell}

\DeclarePairedDelimiter{\abs}{\lvert}{\rvert}

\journal{Medical Image Analysis}

\begin{document}

\begin{frontmatter}


\title{Model Fitting and Analysis of the Discrete Swept Skeletal Representation for Ellipsoidal Objects}

\author[label1]{Mohsen Taheri}
\author[label2]{Stephen M. Pizer}
\author[label1]{J\"orn Schulz}

\affiliation[label1]{organization={Department of Mathematics and Physics, University of Stavanger},
            % city={},
            country={Norway}}

\affiliation[label2]{organization={Department of Computer Science, University of North Carolina at Chapel Hill},
            % city={},
            country={USA}}

%\author{Mohsen Taheri} %% Author name

% \affiliation{organization={Department of Mathematics and Physics, University of Stavanger},%Department and Organization
%             %addressline={}, 
%             %city={},
%             %postcode={}, 
%             %state={},
%             country={Norway}}

%% Abstract
\begin{abstract}
Statistical shape analysis of quasi-ellipsoidal objects, such as groups of hippocampi, is crucial for advancing medical research by aiding in the diagnosis and understanding of various diseases. This work presents a novel object representation method, termed the Locally Parameterized Discrete Swept Skeletal Representation (LPDSSRep). We discuss the model fitting and analysis techniques for this representation, which rely on boundary division and surface flattening. The quality of the model fitting is assessed based on the symmetry and tidiness of the skeletal structure and the volume of the implied boundary. The power of the method is demonstrated by visual inspection and statistical analysis of a synthetic and an actual data set in comparison with an available skeletal representation.
\end{abstract}


% \begin{highlights}
% \end{highlights}

%% Keywords

\end{frontmatter}

\newpage

\section{Introduction}\label{sec:introduction}
A variety of human organs and brain structures have a quasi-ellipsoidal shape, resembling an eccentric ellipsoid\footnote{An eccentric ellipsoid is an ellipsoid with unequal principal radii.}, which we refer to as \textit{Ellipsoidal Objects} (EOs) \citep{Taheri2022}. Thus, the statistical shape analysis of EOs is highly valuable for medical researchers and clinicians. This type of analysis provides crucial insights into identifying differences between groups of brain structures, such as comparing the brain structures of patients with neurodegenerative disorders with those of healthy control subjects \citep{styner2006framework,apostolova2012hippocampal,schulz2013statistical}. The ultimate goal is to assist physicians in diagnosing, predicting, and understanding disorders more accurately, thereby enabling earlier intervention and treatment.\par
The primary need for statistical shape analysis is to establish correspondences among objects within a population based on their geometric properties \citep{laga20193d}. Therefore, this work aims to develop a shape model that facilitates statistical analyses, including hypothesis testing and classification. Our proposed representation establishes correspondence among EOs by partitioning each object into an ordered sequence of slicing planes, where slicing planes with the same relative position correspond across the population.\par
\begin{figure}[ht]
\centering
\boxed{\includegraphics[width=0.98\textwidth]{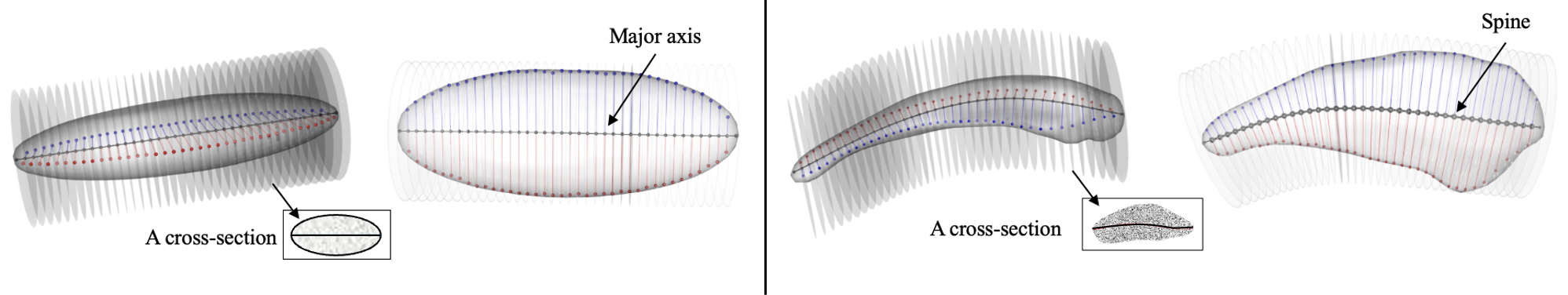}}
\caption{Left: A 3D eccentric ellipsoid showing its major axis and a sequence of slicing planes perpendicular to the axis where each cross-section is an ellipse. Right: A caudate nucleus represented as a 3EO with its spine, such that each cross-section is a 2EO.}
\label{fig:CaudateAsAnEO}
\end{figure}
As illustrated in \Cref{fig:CaudateAsAnEO}, we define a 3EO as a 3D swept region characterized by two vertices\footnote{A vertex is a local relative maximum of convex Gaussian curvature \citep{Pizer2025}.}, an open central curve connecting the vertices, called the \textit{spine}, and a smooth sequence of slicing planes along the spine that do not intersect within the object. Each cross-section (i.e., the intersection of a slicing plane with the object) can be viewed as a 2EO, where a 2EO is a 2D swept region with two vertices connected by a smooth central curve of relatively low curvature throughout. The spine of a 3EO traverses the central curve of each cross-section, ideally through its midpoint. The union of the central curves of the cross-sections forms a relatively flat sheet, which we refer to as the \textit{skeletal sheet} \citep{Taheri2022}. The skeletal sheet can be seen as the 3EO's \textit{skeleton}\footnote{The skeleton of an object is a curve or surface that represents a locally centered abstraction of the shape, obtained through a process of continuous contraction \citep{baerentzen2021skeletonization,siddiqi2008medial}.}. Consequently, the slicing planes simultaneously sweep through both the 3EO's boundary and its skeleton.\par
\begin{figure}[ht]
\centering
\boxed{\includegraphics[width=0.70\textwidth]{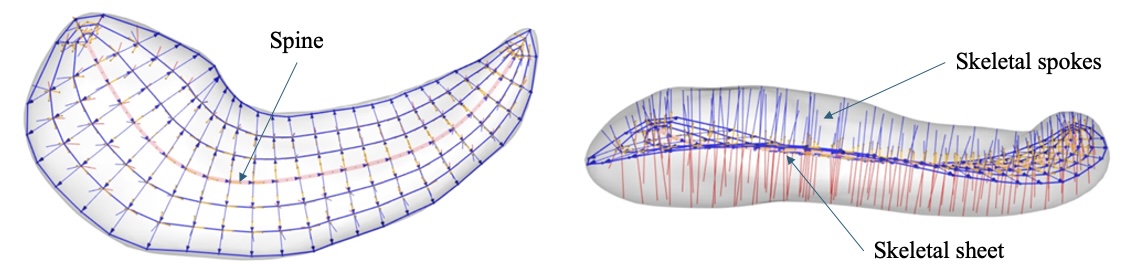}}
\caption{Skeletal structure of a 3EO.}
\label{fig:SkeletalStructure}
\end{figure}
Motivated by \citet{damon2003smoothness,damon2008swept} and \citet{pizer2013nested,pizer2022skeletons}, we understand the \textit{skeletal structure} of a 3EO as a field of radial vectors (i.e., non-crossing internal vectors) that we call \textit{skeletal spokes} with tips on the boundary and tails on the 3EO's skeletal sheet, as depicted in \Cref{fig:SkeletalStructure}. Skeletal spokes provide geometric information such as locational width and direction. A 3EO representation can be considered as the \textit{Discrete Skeletal Representation} (DSRep) as a finite subset of the 3EO's skeletal structure \citep{pizer2013nested,liu2021fitting,Taheri2022}.\par
Following the discussion of \citet{Pizer2025}, meaningful correspondence \citep{van2011survey} across a sample of 3EOs is established through their DSReps under the assumption that 3EOs correspond to an eccentric ellipsoid.\par 
That is, each 3EO has a spine corresponding to the major axis of the ellipsoid and two vertices corresponding to the endpoints of the ellipsoid's major axis. The \textit{crest}\footnote{A crest point is a local relative maximum along one of the principal directions of one of the principal curvatures \citep{Pizer2025}.} (or ridge) of a 3EO is defined as the union of the vertices of its cross-sections and forms a closed curve on the 3EO's boundary. This curve corresponds to the crest of an eccentric ellipsoid, i.e., the intersection of its first principal plane with its boundary (see \Cref{fig:CaudateAsAnEO,fig:SkeletalStructure}).
In this sense, a DSRep is a tuple of skeletal spokes, and a set of DSReps is a set of tuples such that the tuples correspond to each other element-wise. Therefore, we can compare and analyze the corresponding DSReps element-wise \citep{schulz2016non,pizer2020object}.\par
The DSRep has several advantages over most traditional shape representations, such as the \textit{landmark-based model} \citep{dryden1998statistical}, \textit{point distribution model} (PDM) \citep{laga20193d,styner2006framework}, \textit{Euclidean distance matrix} \citep{lele2001invariant}, and \textit{persistent homology} \citep{gamble2010exploring,turner2014persistent}. One key advantage of the DSRep is its ability to capture interior curvature and width along an object \citep{pizer2020object}. Additionally, the DSRep can be parameterized to be invariant to rigid transformations (i.e., alignment-independent), which allows for the accurate detection of local dissimilarities between objects \citep{Taheri2022}. This skeletal model can also explicitly describe different types of dissimilarities, such as shrinkage, bending, and protrusion.\par
Any swept region has a \textit{swept skeletal structure} formed by the union of the skeletal structures of its cross-sections satisfying Damon's \textit{Relative Curvature Condition} (RCC) \citep{damon2008swept}. The RCC establishes a curvature tolerance for the center curve to prevent cross-section intersections within the object, as comprehensively discussed by \citet{TaheriShalmani2026}. Although several deformable medial modeling approaches have been proposed to initialize or fit DSReps to 3EOs \citep{liu2021fitting,Pizer2025,pouch2020automated}, their objectives differ from the swept skeletal modeling considered here. For example, Pouch et al. \citep{pouch2020automated} proposed an automated meshing approach for initializing deformable medial models. To the best of our knowledge, only a few shape models explicitly represent swept skeletal structures, including our recently proposed elliptical tube representation \citep{TaheriShalmani2026}, 
which considers swept regions with elliptical cross-sections under the RCC. Therefore, the objective of this work is to develop a \textit{Discrete Swept Skeletal Representation} (DSSRep) that captures the swept structure of 3EOs and, when fitted to an object, establishes meaningful correspondence across a population of 3EOs.\par
As highlighted in the title of this manuscript, our objective is not to calculate or extract the exact skeletal structure of the target object. Instead, assuming that the target object belongs to the class of 3EOs, we seek a swept skeletal model consistent with this object class whose implied boundary (i.e., the envelope of the skeletal spokes \citep{pizer1999segmentation}) closely approximates the object's boundary (see \Cref{fig:goodnessOfFit}). The resulting model provides a skeletal representation of the target object suitable for statistical shape analysis.\par
One potential method for fitting a DSSRep to a 3EO is to first calculate the spine as a \textit{curve skeleton} \citep{dey2006defining} and then define the cross-sections along the spine. However, there are limitations in using the curve skeleton approach. For example, even a smooth curve skeleton may still have branches, as there is no requirement for a curve skeleton to be a simple curve. There are methods for simplifying the curve skeleton to define a simple curve, such as the Laplacian contraction method of \citet{au2008skeleton}, the mean curvature skeleton of \citet{tagliasacchi2012mean}, the $L1$-medial skeleton of \citet{huang2013l1}, and skeletonization via local separators by \citet{baerentzen2021skeletonization}. However, our observations indicate that these methods fail to provide a suitable spine. Most of these methods indiscriminately search for the 3EO's skeleton without considering key geometric properties, such as the crest and vertices. Besides, these methods typically ignore the RCC, as it is not a requirement for defining the curve skeleton. As a result, the spine can exhibit unpredictable behavior, bending and swinging freely within the object, as discussed in the \supplementary.\par
\begin{figure}[ht]
\centering
\boxed{\includegraphics[width=0.98\textwidth]{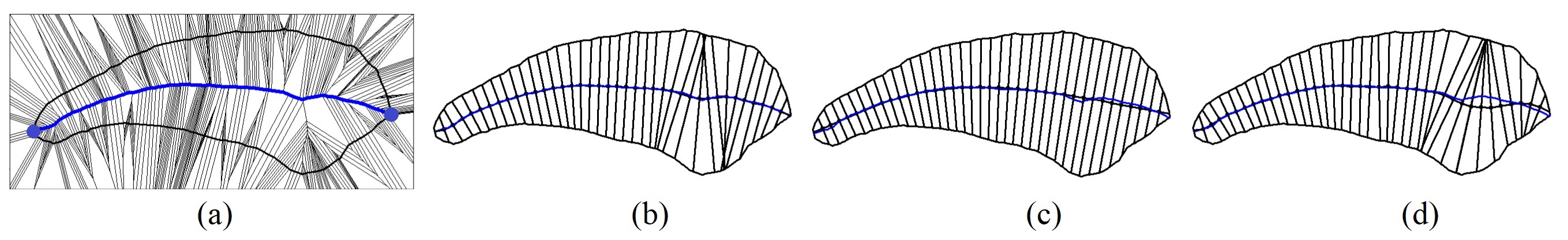}}
\caption{Skeletal structures of a 2EO: (a) Voronoi diagram. The Central Voronoi Skeleton (CVS) is represented by the blue curve connecting the two vertices (blue dots). (b) An invalid model based on a smooth curve near the CVS, which violates the RCC. (c) A valid model based on a slightly relaxed CVS that is neat but not highly symmetric. (d) A valid model with high symmetry, which is close to violating the RCC.}
\label{fig:GC_crossSections}
\end{figure}

Moreover, the swept skeletal structure is not necessarily unique (see \Cref{fig:GC_crossSections}). In fact, as long as the center curve of a swept region satisfies the RCC, it can be deformed while ensuring that the cross-sections do not intersect within the object. Consequently, multiple valid swept skeletal structures may exist, making it necessary to determine which one is the most appropriate one for shape representation.\par
As \Cref{sec:Basic_terms_and_definitions} discusses, an object with a discrete boundary has a unique \textit{Voronoi skeleton} \citep{siddiqi2008medial,attali1997computing,attali2009stability}. Assuming that the boundary of a 3EO can be partitioned properly into two components segmented by the crest, namely the \textit{top part} and the \textit{bottom part}, an EO has a \textit{Central Voronoi Skeleton} (CVS), as \Cref{sec:Central_Voronoi_Skeleton} discusses (see \Cref{fig:caudateSkeletalSurface}). Our approach to obtaining the DSSRep is to use a relaxed version of the CVS to fit the skeleton of the 3EO and then define the skeletal spokes accordingly. To provide a better intuition, consider a 2EO as shown in \Cref{fig:GC_crossSections}.\par
Basically, the Voronoi diagram can be computed for any object with discrete boundary \citep{siddiqi2008medial}, as illustrated in \Cref{fig:GC_crossSections}(a). Also considering the two vertices of the object (bold blue dots), the object can be divided into two parts, as discussed in \Cref{sec:Central_Voronoi_Skeleton}. The CVS of the object is a curve in the middle of the object (shown in blue) that separates the two parts, making it a reasonable candidate for constructing the swept skeletal structure. However, the CVS is not typically smooth and exhibits a bumpy structure, particularly at the branch intersections. In this example, neither the CVS nor any smooth, differentiable curve in its neighborhood can serve as the center curve, as both violate the RCC. The RCC violation is obvious as the cross-sections intersect within the object due to high local curvature, as shown in \Cref{fig:GC_crossSections}(b). As a result, a valid DSSRep cannot be established. Relaxing the CVS allows the center curve to satisfy the RCC, as demonstrated in \Cref{fig:GC_crossSections}(c) and \Cref{fig:GC_crossSections}(d). In \Cref{fig:GC_crossSections}(c), the CVS is slightly relaxed, keeping it close to the original.\par
\Cref{fig:GC_crossSections}(c) shows a valid and relatively tidy model, although it is not perfectly symmetric because the center curve does not pass through the middle of the cross-sections. Further relaxation, as shown in \Cref{fig:GC_crossSections}(d), improves skeletal symmetry but increases the local curvature of the center curve, bringing the model closer to violating the RCC and reducing skeletal tidiness. In general, perfect symmetry with cross-sections normal to the center curve may not be feasible, even for objects with a smooth, non-branching medial skeleton \citep{shani1984splines} (see \Cref{fig:normality_vs_symmetricity} in the \supplementary). This illustrates a trade-off between skeletal symmetry and tidiness, which, together with volume coverage, motivates the goodness-of-fit criteria introduced in \Cref{sec:Goodness_of_fit}. \par
We adapt the same strategy to fit DSSRep of 3EOs by considering that each cross-section is a 2EO containing a smooth sequence of line segments (i.e., 1D cross-sections) along its central curve (see \Cref{fig:GC_crossSections}). These line segments can be viewed as two skeletal spokes extending in opposite directions from a common tail position. The union of the center curves of the cross-sections forms the skeletal sheet, and the union of the skeletal spokes creates the radial vector field on the skeletal sheet.\par
\begin{figure}[ht]
\centering
\boxed{\includegraphics[width=0.98\textwidth]{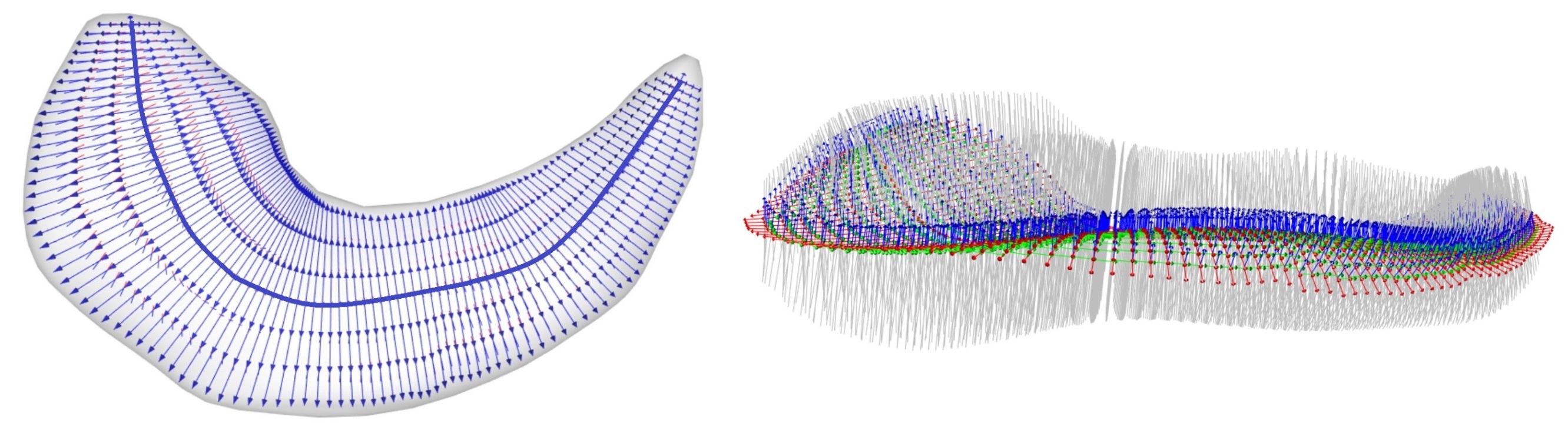}}
\caption{A DSSRep of a hippocampus. Left: Discrete skeletal sheet. Right: Skeletal spokes with tails on the skeletal sheet.}
\label{fig:LP_dss_rep_hippo}
\end{figure}

By defining the correspondence between the spines of different 3EOs based on curve registration \citep{srivastava2016functional}, we can establish a meaningful correspondence across a sample of 3EOs. This provides corresponding spinal cross-sections associated with corresponding points on the spines. Similarly, just as corresponding cross-sections can be defined along the spines, corresponding line segments can be defined along the central curves of the cross-sections using 2D curve registration. This approach is designed to generate a finite set of corresponding skeletal spokes, resulting in the 3EO's DSSRep, as illustrated in \Cref{fig:LP_dss_rep_hippo}.\par
In this work, we use the CVS as the basis for a novel DSSRep model-fitting approach. We define the skeletal sheet by relaxing the 3EO's CVS and subsequently define the spine as a curve on this sheet connecting the two vertices, as discussed in \Cref{sec:Fitting_d_ss_rep}. The goodness of fit is evaluated based on skeletal symmetry, tidiness, and volume coverage. \par
To ensure the analysis is alignment-independent and to effectively capture local dissimilarities (e.g., protrusion, elongation), we extend the concept of the \textit{Locally Parameterized DSRep} (LPDSRep) proposed by \citet{Taheri2022} to introduce the \textit{Locally Parameterized DSSRep} (LPDSSRep). This is achieved by parameterizing the DSSRep using a tree-like structure of its skeletal sheets equipped with local frames, as detailed in \Cref{sec:Parameterization}. The structure of this work is outlined as follows.\par
\Cref{sec:Basic_terms_and_definitions} reviews basic terms and provides explicit definitions of a swept region, 3EO, skeletal structure, and swept skeletal structure. \Cref{sec:Central_Voronoi_Skeleton} introduces the CVS. \Cref{sec:skeletal_structure} presents a DSSRep model fitting procedure based on skeleton flattening using dimensionality reduction methods. \Cref{sec:Parameterization} and \Cref{sec:Goodness_of_fit} introduce the LPDSSRep and discuss the criteria for a proper model based on essential skeletal symmetry, tidiness, and the volume of the implied boundary. \Cref{sec:Statistical_analysis} demonstrates hypothesis testing and classification using LPDSSRep Euclideanization. \Cref{sec:Evaluation} compares LPDSRep and LPDSSRep using a set of toy examples and a real dataset to highlight the strengths and limitations of our method. Finally, \Cref{sec:Conclusion} summarizes the findings and concludes the work.

\section{Basic terms and definitions}\label{sec:Basic_terms_and_definitions}
In this section, we review basic terms and definitions regarding skeletal structures and swept skeletal structures.\par
\subsection{Skeletal structure}
\begin{figure}[ht]
\centering
\boxed{\includegraphics[width=0.98\textwidth]{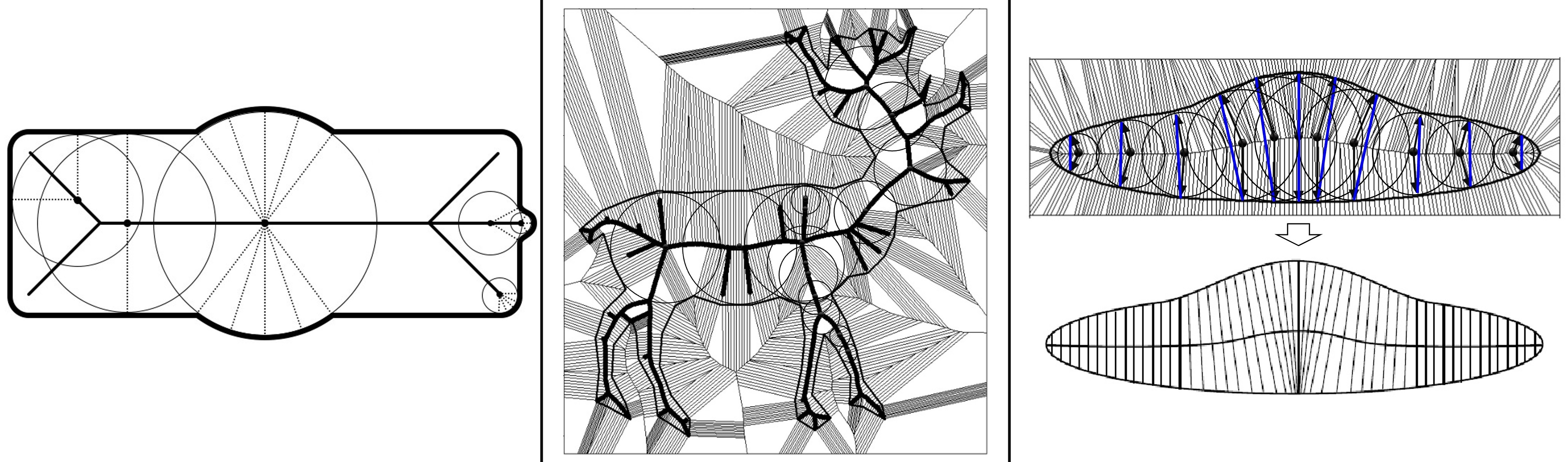}}
\caption{Left: Medial skeleton and maximal spheres of a 2D object. The bold curve is the medial skeleton. The dotted lines are medial spokes. Middle: Voronoi diagram and several maximal inscribed spheres of a 2D deer. The Voronoi skeleton is shown in bold. Right: Chordal skeletal structure a 2D object resembling a 2EO (bottom) obtained from the object's medial skeletal structure (top).}
\label{fig:medial_locus}
\end{figure}
We consider the set $\Omega\subset\mathbb{R}^{d}$ as a $d$-dimensional object if $\Omega$ is homeomorphic to the $d$-dimensional closed ball, where $d\in\mathbb{N}$. We denote the boundary of $\Omega$ by $\partial\Omega$ and its interior by $\Omega_{in}$, so $\Omega=\Omega_{in}\cup\partial\Omega$. Assume point $\bm{p}\in\Omega_{in}$ and a unit direction $\bm{u}\in\mathbb{S}^{d-1}$, where $\mathbb{S}^{d-1}=\{\bm{x}\in\mathbb{R}^{d}\mid\|\bm{x}\|=1\}$ is the unit $(d-1)$-sphere. If we start at $\bm{p}$ and move straight forward along the ray $\bm{p}+\lambda\bm{u}$, $\lambda\geq 0$, we ultimately reach a boundary point. We call such a straight interior path with starting point $\bm{p}$ and direction $\bm{u}$ a \textit{spoke}. We denote a spoke based on its positional and directional components by $\bm{s}_{(\bm{p},\bm{u})}$.\par
We consider the skeletal spokes of $\Omega$ to be the set of all non-crossing spokes emanating from its skeleton $M$, where, in the generic case\footnote{A generic property persists under almost all small changes of the entity \citep{siddiqi2008medial}.}, $M$ is a Whitney stratified set formed by a disjoint union of smooth manifolds (i.e., smooth curves and surfaces) that provides a locally centered abstraction of the object through continuous contraction \citep{damon2003smoothness,baerentzen2021skeletonization}. The skeletal structure of $\Omega$ is a field of skeletal spokes $U$ on $M$ denoted by $(M,U)$ (see \Cref{fig:GC_crossSections,fig:medial_locus}).
\subsubsection{Medial skeletal structure}
The medial skeletal structure is one form of skeletal structure where the object's skeleton is the medial skeleton. The medial skeleton of an object $\Omega$, known as Blum's medial axis \citep{blum1973biological,siddiqi2008medial}, is the set
\begin{equation}\label{equ:medial_locus}
M_\odot=\{\bm{p}\in\Omega_{in}\mid\; \#\{\bm{q}\in\partial\Omega\mid\,\|\bm{p}-\bm{q}\|=d_{min}(\bm{p},\partial\Omega)\}\geq{2}\},    
\end{equation}
where $d_{min}(\bm{p},\partial\Omega)$ is the minimum Euclidean distance between $\bm{p}$ and $\partial\Omega$, and $\|.\|$ and $\#$ represent the Euclidean norm and cardinality, respectively. In other words, $M_\odot$ is the set of centers of all maximal (inscribed) spheres bi-tangent (or multi-tangent) to $\partial\Omega$. A medial spoke is a spoke connecting the center of a maximal sphere to its tangency point, normal to the boundary. The collection of all medial spokes $U_\odot$ on the medial skeleton $M_\odot$ forms the medial skeletal structure $(M_\odot,U_\odot)$ \citep{siddiqi2008medial}, where 
\[
U_\odot=\{\bm{s}_{(\bm{p},\bm{u})}\mid \bm{p}\in{M}_\odot \;\text{and}\; \|\bm{s}_{(\bm{p},\bm{u})}\|=d_{min}(\bm{p},\partial\Omega)\}.
\]
\Cref{fig:medial_locus} (left) illustrates the medial skeleton and a few medial spokes.\par
\subsubsection{Voronoi skeleton}
For an object with a discrete boundary represented by a finite set of boundary points, the \textit{Voronoi skeleton} is defined using the Voronoi diagram of the boundary samples. Let $P=\{\bm{p}_1,\ldots,\bm{p}_n\}\subset\partial\Omega$ denote the set of boundary sample points. The \textit{Voronoi cell} associated with a boundary point $\bm{p}_i$ is defined as $\left\{\bm{x}\in\mathbb{R}^d\;\middle|\;\|\bm{x}-\bm{p}_i\|\le\|\bm{x}-\bm{p}_j\|,\ \forall j\neq i\right\}$.
A point belongs to the \textit{Voronoi skeleton} if it lies on the intersection of two or more Voronoi cells and is contained within the object, as shown in \Cref{fig:medial_locus} (Middle). Given a sufficiently dense sampling of the boundary, the Voronoi skeleton converges to the medial skeleton \citep{dey2004approximating,siddiqi2008medial,attali2009stability}.\par

\subsection{Swept skeletal structure}\label{sec:BasicTermsSweptskeletalstructure}
A swept region is a $d$-dimensional object with a smooth sequence of slicing planes along a center curve (not necessarily normal to the center curve) such that cross-sections do not intersect within the object, and each cross-section is a $(d-1)$-dimensional object \citep{damon2008swept}. In other words, the slicing planes sweep the object's interior and boundary (see \Cref{fig:CaudateAsAnEO,fig:SkeletalStructure,fig:GC_crossSections}).\par
Let $\Gamma$ be the center curve of the swept region $\Omega$ with skeleton $M$. Assume $\Gamma_{(t)}$ as the curve length parameterization of $\Gamma$ such that $t\in[0,1]$, where $\Gamma_{(0,1)}$ denotes the curves' endpoints \citep{srivastava2016functional}. Let $\Pi_{(t)}$ be the slicing plane crossing $\Gamma_{(t)}$, and let $\Omega_{(t)}=\Pi_{(t)}\cap\Omega$ be the cross-section at $\Gamma_{(t)}$. Let $U_{(\bm{p})}$ denote the set of skeletal spokes with tails on $\bm{p}\in{M}$. The skeletal structure of $\Omega$ is a swept skeletal structure if, for each $\bm{p}\in{\Omega_{(t)}}\cap{M}$, the $U_{(\bm{p})}\in\Omega_{(t)}$, i.e., all the skeletal spokes with tails on a cross-section are coplanar. Thus, the slicing planes also sweep the object's skeletal structure \citep{damon2008swept}.\par
As depicted in \Cref{fig:medial_locus} (right),  if the medial skeleton $M_\odot$ of a 2D object can be seen as an open curve, then for each medial point $\bm{p}$ there are two spokes namely up and down spokes on two sides of the medial skeleton with the tail at $\bm{p}$ and the tip at the object boundary as
\begin{equation}\label{equ:boundaryPoints}
\bm{b}=\bm{p}-\mathscr{R}(l)\abs{\frac{d}{dl}{\mathscr{R}(l)}}\bm{t}\pm{\mathscr{R}(l)}\sqrt{1-\abs{\frac{d}{dl}{\mathscr{R}(l)}}^2}\bm{n},
\end{equation}
where $\bm{n}$ and $\bm{t}$ are normal and tangent vectors of the medial skeleton at $\bm{p}$, and $\mathscr{R}$ is the radius function based on curve length parameterization $l$ such that $\mathscr{R}(l)$ is the radius of the maximal inscribe sphere centered at $\bm{p}$ \citep{giblin2003intrinsic}.\par
By connecting the tips of the up and down spokes with the same tail positions, we obtain a set of chords as non-crossing line segments, defining a \textit{chordal skeletal structure}, where the skeleton is a curve as the set of the midpoints of all chords, as discussed by \citet{giblin1985local}. The chordal skeletal structure is a form of swept skeletal structure, as discussed by \citet{damon2008swept}. In this sense, an ellipse is a 2EO with a chordal skeletal structure, where the skeleton is the major axis and the slicing planes are perpendicular to the skeleton \citep{brady1984smoothed,giblin1985local,TaheriShalmani2026}.\par
As mentioned in \Cref{sec:introduction}, we consider a 3EO as a swept region with a swept skeletal structure such that each cross-section $\Omega(t)$ is a 2EO and the length of the spine $\Gamma$ (i.e., the 3EO's center curve) is notably larger than the length of the center curve of each cross-section. The intersection of the spine with each 2EO is a point on and approximately at the middle of the 2EO's skeleton. The union of the 2EOs' boundaries forms the 3EO's boundary, and the union of the 2EOs' skeletons forms the 3EO's skeleton, called the skeletal sheet. Thus, any eccentric ellipsoid is a 3EO by considering the ellipsoid's major axis as the spine and (parallel) slicing planes perpendicular to the spine with cross-sections as ellipses, which are 2EOs.

\section{Central Voronoi skeleton}\label{sec:Central_Voronoi_Skeleton}
In this section, we discuss the Central Voronoi skeleton (CVS) of 2EOs and 3EOs as a subset of their Voronoi skeleton. In \Cref{sec:skeletal_structure}, we use the CVS to fit the 3EO's DSSRep.\par
\begin{figure}[ht]
\centering
\boxed{\includegraphics[width=0.67\textwidth]{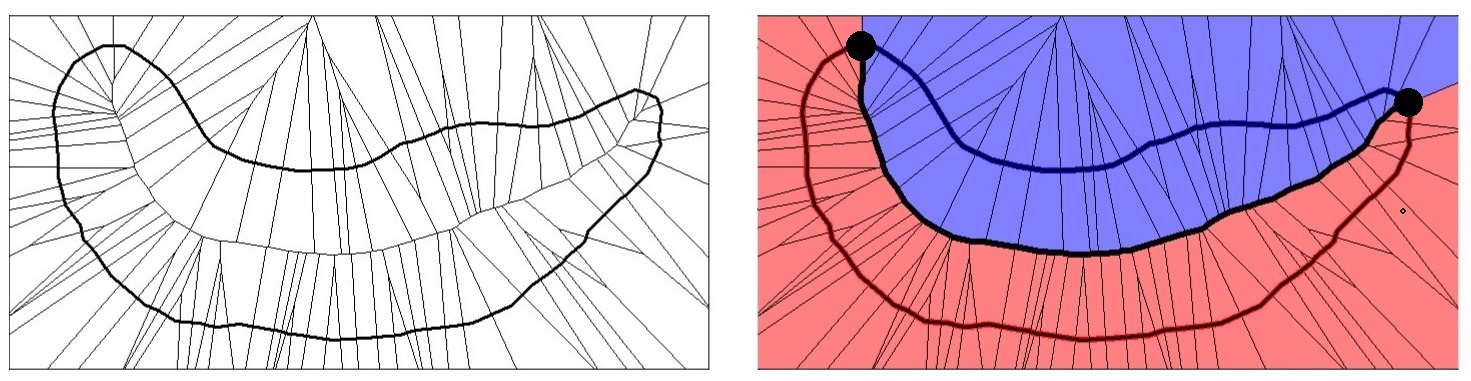}}
\caption{Left: The Voronoi diagram of a 2EO. Right: The top and bottom parts are depicted as blue and red curves. The two sub-regions are two union polygons associated with the top and bottom parts, with the same color. The CVS is the black curve as part of the shared boundary of the union polygons connecting the object vertices.}
\label{fig:Voronoi2D}
\end{figure}
Assume $\Omega$ as a 2EO object with two vertices. The two vertices divide the boundary $\partial\Omega$ into two non-overlapping components, namely the top part and the bottom part. Each part is a simply connected manifold with no holes or discontinuities. These two parts cover the entire boundary without any gaps. In a discrete sense, $\partial\Omega$ and its two parts can be represented by a set of adjacent points. Thus, each boundary point belongs to only one part, and each part is a set of adjacent points \citep{de2000computational}. Let $\partial\hat{\Omega}$ be the discrete form of $\partial\Omega$, and $\partial\hat\Omega^+$ and $\partial\hat\Omega^-$ be the top and bottom parts, respectively. Thus, $\partial\hat\Omega=\cup\partial\hat\Omega^\pm$. Assume a box that contains $\partial\hat{\Omega}$. The Voronoi diagram of each part consists of a set of adjacent polygons such that adjacent polygons share a common edge. The union of these adjacent polygons is a union polygon as a connected subset of the box. Therefore, the box (or the embedding space) is partitioned into two sub-regions. Let $\Omega^+$ and $\Omega^-$ be the intersection of the two sub-regions with $\Omega$ associated with $\partial\hat\Omega^+$ and $\partial\hat\Omega^-$. Thus, $\Omega^+$ and $\Omega^-$ can be seen as two sub-objects, such that $\Omega=\cup\Omega^\pm$. The intersection of these sub-objects $\Omega^+\cap\Omega^-$ defines their shared boundary, which we consider as the CVS of $\Omega$.\par
Each point on the CVS is equidistant from the top and bottom boundary parts because it belongs to the intersection of Voronoi cells associated with the top and bottom boundary sample points. Thus, the CVS is a unique subset of the Voronoi skeleton that is central with respect to $\partial\hat\Omega^+$ and $\partial\hat\Omega^-$. In this sense, the CVS can be seen as the set of centers of maximal spheres that are bitangent to both top and bottom boundary components, as illustrated in \Cref{fig:CSV_vs_Medial} of \supplementary.\par
Further, the CVS has no branches because it is part of $\partial\Omega^+$ (i.e., the boundary of the object $\Omega^+$). Also, the CVS has no discontinuity because if it has a discontinuity, it means $\Omega^+$ and $\Omega^-$ are adjacent but there is a gap between them. Thus, there is a polygon inside $\Omega$ whose interior does not belong to $\Omega^+$ or $\Omega^-$, which contradicts the definition of the Voronoi diagram. \Cref{fig:Voronoi2D} illustrates the Voronoi diagram and the CVS of a 2EO.\par
\begin{figure}[ht]
\centering
\boxed{\includegraphics[width=0.97\textwidth]{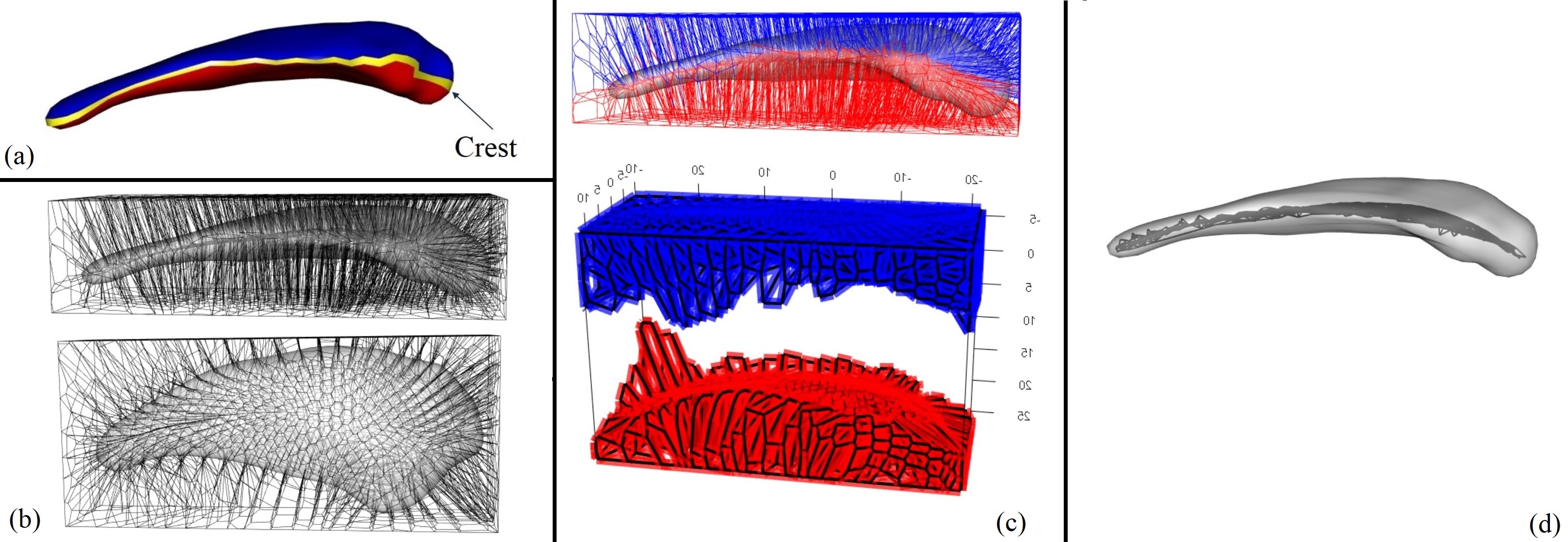}}
\caption{(a) The top part, bottom part, and crest of a caudate are shown in blue, red, and yellow, respectively. (b) Voronoi diagram of the caudate from two different perspectives. (c) Top: Illustration of the two sub-regions in blue and red. Bottom: The sub-regions are separated to enhance visualization. (d) The CVS is represented as a black triangular surface. }
\label{fig:caudateSkeletalSurface}
\end{figure}
Similarly, by assuming the crest of a 3EO as a closed curve (corresponding to the crest of an eccentric ellipsoid), the crest divides the 3EO's boundary into two parts, the top and bottom parts. The Voronoi diagrams of the two parts define two union polyhedra (i.e., two 3D polygons) as two sub-objects. The CVS of a 3EO is the shared boundary of these two sub-objects. Thus, the CVS is a unique subset of the medial skeleton. Analogous to the CVS of a 2EO, the CVS of a 3EO is a manifold with no holes or branches. \Cref{fig:caudateSkeletalSurface} (a) illustrates the top part, the bottom part, and the crest of a caudate nucleus as a 3EO. \Cref{fig:caudateSkeletalSurface} (b) shows the Voronoi diagram of the 3EO located in a box. \Cref{fig:caudateSkeletalSurface} (c) depicts the two sub-regions associated with the top and bottom parts. The sub-regions are also separated to provide a clear intuition about the sub-regions. \Cref{fig:caudateSkeletalSurface} (d) illustrates the CVS.

\section{Fitting swept skeletal structure}\label{sec:skeletal_structure}
In this section, we discuss fitting swept skeletal structures for 2EOs, and then we generalize the method for 3EOs.\par
We assume the skeleton, or the center curve of a 2EO, is a smooth open curve. Along the boundary of a 2EO, the normal directions generally do not change abruptly except near the vertices associated with the endpoints of the center curve, where the curvature reaches local maxima \citep{siddiqi2008medial,Pizer2025}. The same discussion is valid for a 3EO near the crest \citep{giblin2003intrinsic,giblin2004formal,siddiqi2008medial,fletcher2004principal}. \citet{abulnaga2021volumetric} used this property to calculate the crest and the two boundary parts. They considered the discrete boundary and measured the geodesic distance between any two boundary points (to find how far apart they are) and their normal difference. Then, they applied spectral clustering of \citet{ng2001spectral} (as a binary classification method) to classify the boundary points into the top and bottom parts.\par
The main idea of our model fitting is to apply Abulnaga's method to divide the 3EO's boundary, obtain the CVS based on the two boundary parts, fit a smooth manifold representing the skeletal sheet close to the CVS, and finally find the spokes from the skeleton to the boundary such that the obtained skeletal structure is a swept skeletal structure. To provide intuition, we begin by considering a 2EO, as depicted in \Cref{fig:toyExample}.

\subsection{Fitting a swept skeletal structure to a 2EO}\label{sec:2DTubeSkeletalFitting}
\begin{figure}[ht]
\centering
\boxed{\includegraphics[width=0.97\textwidth]{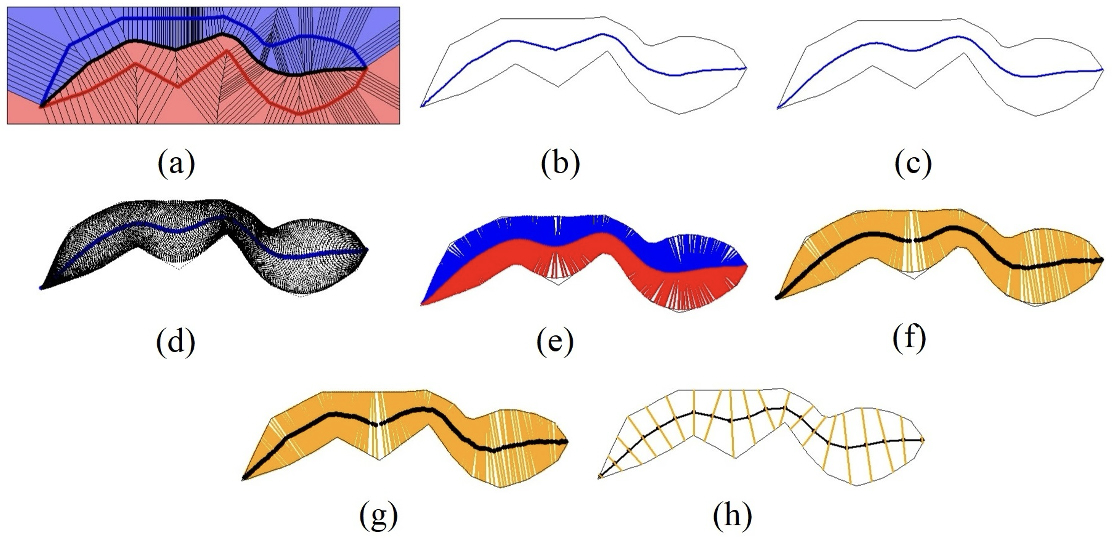}}
\caption{Obtaining the center curve of a 2EO. (a) Blue and red show the sub-regions associated with the boundary's top and bottom parts. (b) The CVS. (c) The relaxed CVS. (d) The maximal inscribed circles are centered at the relaxed CVS. The implied boundary is the envelope of the inscribed circles. (e) Up and down skeletal spokes of a large number of points on the relaxed CVS in blue and red. (f) Chordal structure of the implied boundary. (g) Quasi-chordal structure of the object. Black dots show the quasi-chordal skeleton. (h) The DSSRep of the 2EO based on the curve length registration.}
\label{fig:toyExample}
\end{figure}
Let $\Omega_2$ be a 2EO represented by an elongated and slightly deformed ellipse. Our objective is to fit a DSSRep by constructing a smooth center curve $\Gamma_{\Omega_2}$ connecting the two vertices and a sequence of slicing lines (i.e., 1D planes) normal to this curve. Let $\Gamma_{\Omega_2}^0$ and $\Gamma_{\Omega_2}^1$ denote the endpoints of the center curve, corresponding to the two vertices of $\Omega_2$. Furthermore, let $\partial\Omega_2^+$ and $\partial\Omega_2^-$ denote the top and bottom parts of $\partial\Omega_2$, respectively.\par
Thus, $\Gamma_{\Omega_2}^{0,1}$ are at the edge of the top and bottom parts. Therefore, to find the $\Gamma_{\Omega_2}^{0,1}$ and the CVS of the object, it is sufficient to find the top and bottom parts.\par
To identify the top and bottom boundary parts, we apply the spectral clustering approach of \citet{abulnaga2021volumetric}. Based on the assumption that boundary-normal directions change substantially only near the vertices $\Gamma_{\Omega_2}^\pm$, the affinity matrix is defined as
$[W]_{i,j}=\exp\{\delta\langle\bm{n}_i,\bm{n}_j\rangle{d_g(\bm{x}_i,\bm{x}_j)}\}$, where $\bm{n}_i$ is the normal vector at the $i$th boundary vertex $\bm{x}_i$, $d_g$ is the geodesics distance, and $\delta$ controls the local effect of normal variation. Therefore, the elements of the affinity matrix reflect the distance between two boundary points and how much their normals are different. Based on spectral clustering of \citet{ng2001spectral}, the sign of the second smallest eigenvectors of the normalized Laplacian $\mathcal{L}=I-D^\frac{-1}{2}WD^\frac{-1}{2}$ classifies the boundary points into two classes as top and bottom parts that are consistent with the orientation of the normals, where $I$ is the identity matrix, and $[D]_{i,j}=\Sigma_j{[W]_{i,j}}$.\par
We also verify that the boundary parts are connected, i.e., the geodesic between any two points within a boundary part remains within that part, ensuring spatially coherent boundary clusters. If this condition is violated, $\delta$ is adjusted heuristically until a valid partition is obtained. We postpone the development of an automatic adjustment procedure to future work. The same discussion is valid for 3EOs (see \Cref{fig:ellipsoidal}). In this work, we choose $\delta=0.5$, which typically produces a reasonable crest; its effect is illustrated for the caudate in \supplementary. \par
As discussed in \Cref{sec:Central_Voronoi_Skeleton}, by having the top and bottom parts, we can calculate the CVS based on the Voronoi diagram, as shown in \Cref{fig:toyExample} (a). The two sub-regions are in blue and red. \Cref{fig:toyExample} (b) depicts the CVS, which is not smooth. We relax the CVS by fitting a smooth curve close to it (e.g., by implicit polynomial \citep{unsalan1999new}, principal curve \citep{hastie1989principal}, nonlinear regression \citep{ritz2008nonlinear}, etc.), as depicted in \Cref{fig:toyExample} (c). Let $\hat{\Gamma}_{\Omega_2}$ be the relaxed CVS. Considering $\hat{\Gamma}_{\Omega_2}$, as a dense discrete curve, we generate maximal inscribed circles centered at the points of $\hat{\Gamma}_{\Omega_2}$ as illustrated in \Cref{fig:toyExample} (d). We consider the envelope of the maximal inscribed spheres as the boundary of an object $\hat{\Omega}_2$ such that $\hat{\Gamma}_{\Omega_2}$ represents the medial skeleton of $\hat{\Omega}_2$.\par
We know that when the medial skeleton is an open curve, we can produce the chordal structure, as for each medial point $\bm{p}$ there are two spokes on two sides of the medial skeleton with the tail at $\bm{p}$ and the tip at the object boundary, as discussed in \Cref{sec:BasicTermsSweptskeletalstructure}. Therefore, for each medial point, we generate two spokes pointing toward the top and bottom parts as depicted in \Cref{fig:toyExample} (e). Since there is no guarantee that the spokes do not intersect, for example, when the local curvature of $\hat{\Gamma}_{\Omega_2}$ violates the RCC, we detect intersecting spokes and shorten both the intersecting spokes and their corresponding twin spokes (i.e., the mirror spokes sharing the same tail point). Note that we can also make the relaxed CVS curvier as long as it produces non-intersecting spokes (similar to \Cref{fig:GC_crossSections} (d)).\par
We consider two top and bottom parts for $\partial\hat{\Omega}_2^\pm$ based on the endpoints of $\hat{\Gamma}_{\Omega_2}$. We call spokes connecting medial points to the $\partial\hat{\Omega}_2^+$ and $\partial\hat{\Omega}_2^-$ as \textit{up} and \textit{down} spokes, respectively. The relaxed CVS equipped with the spokes as shown in \Cref{fig:toyExample} (e) can be seen as the quasi-medial skeletal structure of the 2EO if the implied boundary of the spokes approximates the object boundary \citep{pizer1999segmentation,fletcher2004principal}. In other words, $\hat\Omega_2$ represents $\Omega_2$ if the Jaccard index $J(\Omega_2,\hat\Omega_2)\approx{1}$, where $J(\Omega_2,\hat\Omega_2)=\frac{\|\Omega_2\cap\hat\Omega_2\|_A}{\|\Omega_2\cup\hat\Omega_2\|_A}$ and $\|.\|_A$ measures the area. To approximate the Jaccard index between any two objects, we apply a Monte Carlo point-based estimation. We consider a common bounding box enclosing both objects. A large number of points are then sampled uniformly from the bounding box. The Jaccard index is estimated as the ratio of the number of sampled points lying inside both objects to the number of sampled points inside at least one of the two objects \citep{covre2022monte}. In \Cref{sec:volume_Coverage}, we discuss the Jaccard index based on the volume coverage for our 3D models.\par
Connecting the tips of paired up and down spokes yields the \textit{chordal skeletal structure} of $\Omega_2$ (\Cref{fig:toyExample} (f)) as a set of non-crossing line segments connecting $\partial\hat{\Omega}_2^+$ and $\partial\hat{\Omega}_2^-$, as explained in \Cref{sec:BasicTermsSweptskeletalstructure}. We then extend the chords until they reach the target boundary or the RCC threshold to obtain \textit{quasi-chords} (\Cref{fig:toyExample}(g)). For a 2D swept region, the RCC is defined by $r < \frac{1}{\kappa}$, where $r$ denotes the object's width measured along the normal direction and $\kappa$ is the curvature of the central curve \citep{TaheriShalmani2026}.\par
The quasi-chords represent the cross-sections of the 2EO $\Omega_2$, and the curve connecting the middle of the quasi-chords, called the \textit{quasi-chordal skeleton}, represents the center curve of $\Omega_2$. If quasi-chords intersect between the implied boundary $\partial\hat\Omega_2$ and the target boundary $\partial\Omega_2$, they are trimmed at the intersection. This step can be omitted when the relaxed CVS has sufficiently low curvature (because based on the RCC, $r\rightarrow\infty$ when $\kappa\rightarrow 0$). The resulting \textit{quasi-chordal structure} (i.e., quasi-chords plus the quasi-chordal skeleton) therefore provides cross-sections of the target 2EO while satisfying the RCC. The quasi-chordal structure can also be used for straightening the 2EO, as discussed in \supplementary. Finally, curve registration, primarily based on relative arc-length positions as discussed in \citet{srivastava2016functional}, is used to establish correspondence between quasi-chords across a sample of 2EOs, yielding the DSSRep, as shown in \Cref{fig:toyExample}(h).

\subsection{Fitting a swept skeletal structure to a 3EO}\label{sec:Fitting_d_ss_rep}
\begin{figure}[ht]
\centering
\boxed{\includegraphics[width=0.97\textwidth]{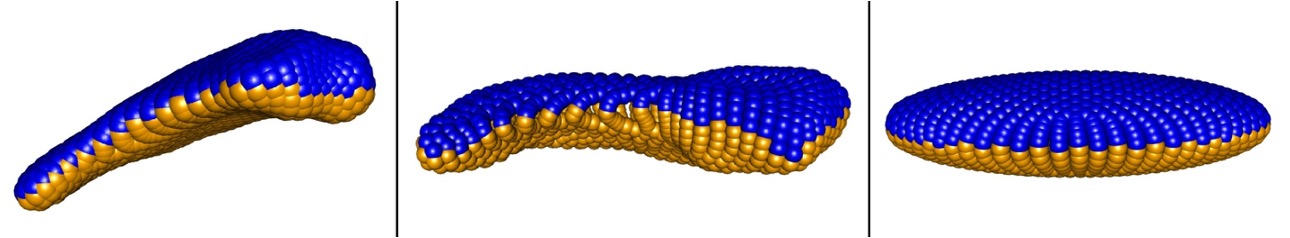}}
\caption{Visualization of 3EOs. Blue and yellow indicate the top and bottom parts of a caudate, a hippocampus, and an ellipsoid from left to right.}
\label{fig:ellipsoidal}
\end{figure}

\begin{figure}[ht]
\centering
\boxed{\includegraphics[width=0.97\textwidth]{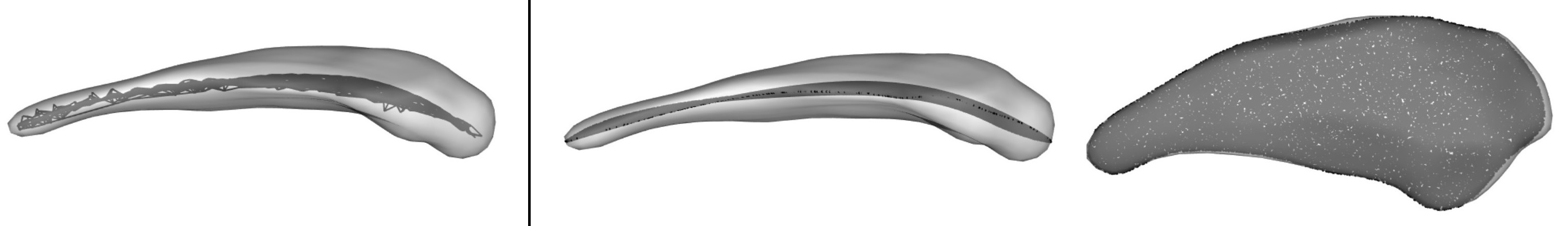}}
\caption{Left: The CVS of a caudate. Right: Two views of the relaxed CVS.}
\label{fig:caudateSkeletalSurface2}
\end{figure}

For fitting a DSSRep to a 3EO, similar to fitting a DSSRep to a 2EO, we divide $\partial\Omega_3$ into the top and bottom parts $\partial\Omega_3^\pm$. \Cref{fig:ellipsoidal} shows the top and bottom parts of a caudate nucleus, a hippocampus, and an ellipsoid.
We consider the border between $\partial\Omega_3^+$ and $\partial\Omega_3^-$ as the crest denoted by $\partial\Omega_3^0$ as shown in \Cref{fig:caudateSkeletalSurface} (a). Thus, $\partial\Omega_3^0$ corresponds to the crest of an eccentric ellipsoid, which is an ellipse, i.e., the intersection of the ellipsoid's first principal plane with its boundary. Based on $\partial\Omega_3^\pm$, we obtain the CVS, as depicted in \Cref{fig:caudateSkeletalSurface} (d) and \Cref{fig:caudateSkeletalSurface2} (left). We relax the CVS by fitting a smooth surface close to it (e.g., based on spline surface fitting \citep{lee1997scattered}).  We consider the relaxed CVS as the skeletal sheet, as visualized in \Cref{fig:caudateSkeletalSurface2} (right).

Based on our observations, the skeletal sheet of various brain subcortical structures like the hippocampus and caudate nucleus has low curvature everywhere, and its flattened version can be seen as a 2EO with two vertices. We call a 3EO with such a skeletal sheet a \textit{regular 3EO}; otherwise, an \textit{irregular 3EO}.\par
Note that any \textit{elliptical tube}\footnote{An elliptical tube is a 3D swept region whose cross-sections are elliptical discs, with the skeletal sheet defined as the union of the cross-sections' major axes \citep{TaheriShalmani2026}.} as defined by \citet{TaheriShalmani2026} can be considered as an irregular 3EO, for which model fitting may be extremely challenging and is beyond the scope of this work. Therefore, we focus on regular 3EOs, while irregular 3EOs are briefly discussed in the \supplementary\ using a mandible without coronoid processes as an example. \par
For fitting the spine (and consequently slicing planes) of a regular 3EO, the idea is to flatten the skeletal sheet based on a suitable manifold dimensionality reduction method to become a 2EO. Then, we fit the center curve of the 2EO, map the fitted center curve to the skeletal sheet, and consider it as the 3EO's spine. In this sense, the spine is a curve on the skeletal sheet connecting two vertices of the 3EO corresponding to the curve connecting the two vertices of the flattened skeletal sheet.

\begin{figure}[ht]
\centering
\boxed{\includegraphics[width=0.97\textwidth]{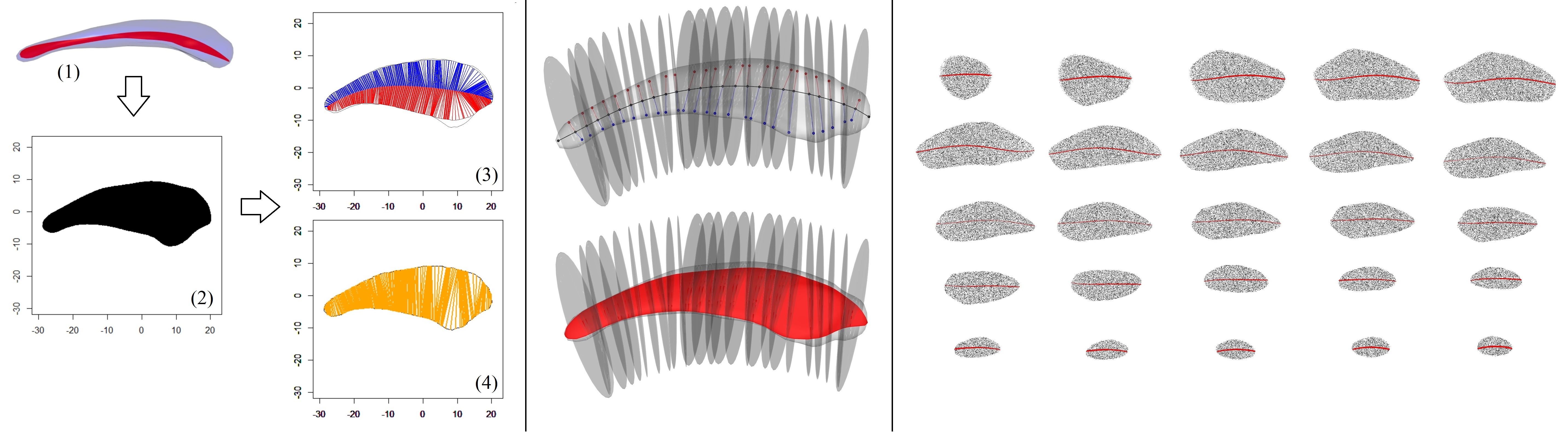}}
\caption{Left: (1) Skeletal sheet of a caudate in red. (2) The PCA projection of the skeletal sheet as a 2EO. (3) Relaxed CVS and the implied boundary of the 2EO. (4) Quasi-chordal structure of the 2EO. Middle: Visualization of the slicing planes along the spine of a caudate (top) plus the skeletal sheet of the object (bottom). Right: cross-sections and their center curves as the intersection of the slicing planes with the skeletal sheet.}
\label{fig:caudate_cross_sections}
\end{figure}
An appropriate dimensionality reduction method preserves the structure of the high-dimensional data properly in the mapped low-dimensional data \citep{van2008visualizing}. For example, \textit{Principal Component Analysis} (PCA) \citep{hotelling1933analysis} is a suitable method when the data has an approximately flat structure relative to the first PCA principal plane (i.e., a plane that is spanned by the first and second eigenvectors with origin at the data centroid). In other words, the flattened version of a 3D surface (flattened by PCA) should not be significantly different from its original version. There are various ways to quantify the irregularity or non-flatness of a surface \citep{bosche2014automating,Haitjema2017,miko2021assessment}. In this work, we quantify the irregularity based on the maximum local curvature. \par
For an entirely flat 2D surface, the absolute values of the two principal curvatures (i.e., the eigenvalues of the second fundamental form) are zero everywhere \citep{pressley2013elementary}. Assume $M$ as a 2D surface. There is a point on $M$ with absolute average curvature $\kappa_{max}\in[0,\infty)$ such that $\forall\bm{p}\in M$; $\kappa_{\bm{p}}\leq\kappa_{max}$, where $\kappa_{\bm{p}}$ is the average absolute value of the two principal curvatures at $\bm{p}$. Therefore, $\kappa_{max}$ can be assumed to be the total curvature of the surface. Thus, the irregularity of $M$ can be strictly quantified as ${2\arctan(\kappa_{max})/\pi\in[0,1]}$. Obviously, $M$ is entirely flat if it has zero irregularity.\par
Let $\tilde{M}_{\Omega_3}$ be the skeletal sheet of $\Omega_3$. In this work, we say $\tilde{M}_{\Omega_3}$ is \textit{quasi-flat} if its irregularity is less than $0.01$. Let $\tilde{M}_{\Omega_3}$ be quasi-flat and assume the mapping ${\mathscr{F}:\tilde{M}_{\Omega_3}\rightarrow\tilde{M}'_{\Omega_3}}$ as the orthogonal projection that maps $\tilde{M}_{\Omega_3}$ to the PCA first principal plane along the third principal axis. We say the quasi-flat $\tilde{M}_{\Omega_3}$ is \textit{PCA flatable} if $\mathscr{F}$ is a diffeomorphism (i.e., $\mathscr{F}$ is a topological preservative bijective mapping between $\tilde{M}_{\Omega_3}$ and $\tilde{M}'_{\Omega_3}$). We consider $\Omega_3$ as a regular 3EO if $\tilde{M}_{\Omega_3}$ is quasi-flat, PCA flatable, and $\tilde{M}'_{\Omega_3}$ can be seen as a 2EO.\par
Let $\Omega_3$ be a regular 3EO, and let $\Gamma'_2$ be the center curve of the embedded 2EO $\tilde{M}'_{\Omega_3}$. We consider $\mathscr{F}^{-1}(\Gamma'_2)$ as the spine of $\Omega_3$. Also, if $c$ is a quasi-chord of $\tilde{M}'_{\Omega_3}$ we consider $\mathscr{F}^{-1}(c)$ as a non-linear quasi-chord of $\tilde{M}_{\Omega_3}$. Note that since we apply orthogonal projection, both $c$ and $\mathscr{F}^{-1}(c)$ are located on a plane. Therefore, we can consider these planes as the slicing planes of $\tilde{M}_{\Omega_3}$ and also the slicing planes of $\Omega_3$, as far as conditions such as the RCC in 3D are satisfied. We leave a detailed investigation of these conditions for future research. \Cref{fig:caudate_cross_sections} shows the cross-sections of a caudate obtained based on the chordal structure of the PCA projection of the skeletal sheet.\par
\begin{figure}[ht]
\centering
\boxed{\includegraphics[width=0.70\textwidth]{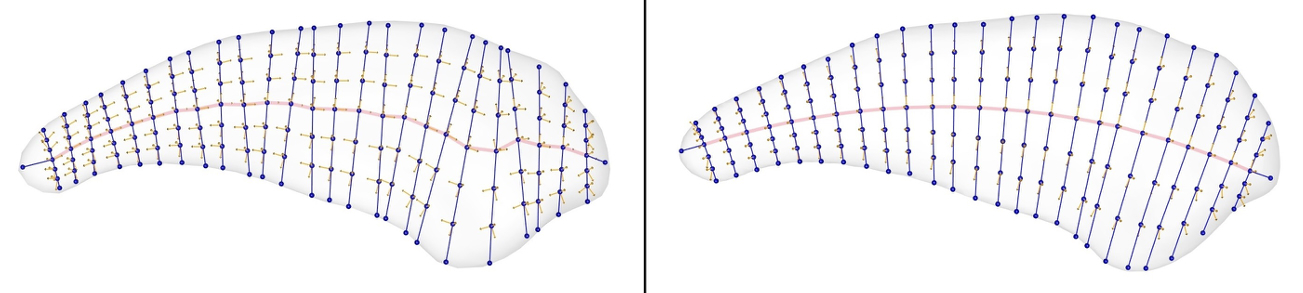}}
\caption{Fitted tree-like structure of the skeletal sheet of a caudate. Left: Slicing planes and the spine are based on chordal structure. Right: The spine is based on the relaxed CVS, and the slicing planes are normal to the spine.}
\label{fig:caudateHippoLocalFrames}
\end{figure}

As depicted in \Cref{fig:caudateHippoLocalFrames}, the quasi-chordal skeleton (i.e., the union of the midpoints of the quasi-chords) is typically more wavy than the CVS, which is undesirable because it reduces the \textit{skeleton tidiness}, as discussed in \Cref{sec:SkeletonPerturbation}. As stated in \Cref{fig:caudateHippoLocalFrames}, the objective of this work is not to calculate or extract the exact skeletal structure of the target object. Instead, we fit candidate swept skeletal models consistent with the 3EO class and select the one that best represents the target object.
In this sense, the $\Gamma'_2$ can be considered as the relaxed CVS of $\tilde{M}'_{\Omega_3}$, which is also beneficial for avoiding violations of the RCC in the resulting spine $\mathscr{F}^{-1}(\Gamma'_2)$, as it has relatively lower curvature throughout. Further, we can define the cross-sections to be perpendicular to the spine. This construction is also more consistent with the definition of the moving Frenet frame along the spine \citep{TaheriShalmani2026}. \Cref{fig:caudateHippoLocalFrames} illustrates the spine obtained from the quasi-chordal skeleton (left) and the relaxed CVS with cross-sections normal to the spine (right).

\section{Parameterization}\label{sec:Parameterization}
As \citet{lele2001invariant} discussed, it is crucial for statistical shape analysis that the shape representation is invariant to rigid transformations, as shape analysis based on \textit{noninvariant} shape representation is alignment-dependent, while the act of alignment makes the analysis biased and misleading. For example, using a noninvariant shape representation (e.g., SPHARM-PDM) for hypothesis testing of local shape differences between two groups can result in a large number of false positives and false negatives \citep{Taheri2022}.\par
The DSRep can be parameterized to obtain an invariant shape representation that also characterizes the type of local shape dissimilarity. To this end, \citet{Taheri2022} introduced local frames for DSReps and proposed the \textit{Locally Parameterized DSRep} (LPDSRep), an invariant skeletal shape representation based on a tree-like structure of the skeletal sheet.\par
The tree-like structure of the LPDSRep consists of the spine and a set of non-intersecting curves on the skeletal sheet, called \textit{veins}, that connect the spine to the 3EO's crest (analogous to \Cref{fig:skeletal_sheet}). The spine and veins define paths for moving frames on the skeletal sheet. Using the center curves of the non-intersecting cross-sections as veins, we extend this structure to the DSSRep and, following \citet{Taheri2022}, introduce an invariant shape representation called \textit{Locally Parameterized DSSRep} (LPDSSRep).

\subsection{LPDSSRep}\label{sec:LP_dss_rep}
\begin{figure}[ht]
\centering
\boxed{\includegraphics[width=0.97\textwidth]{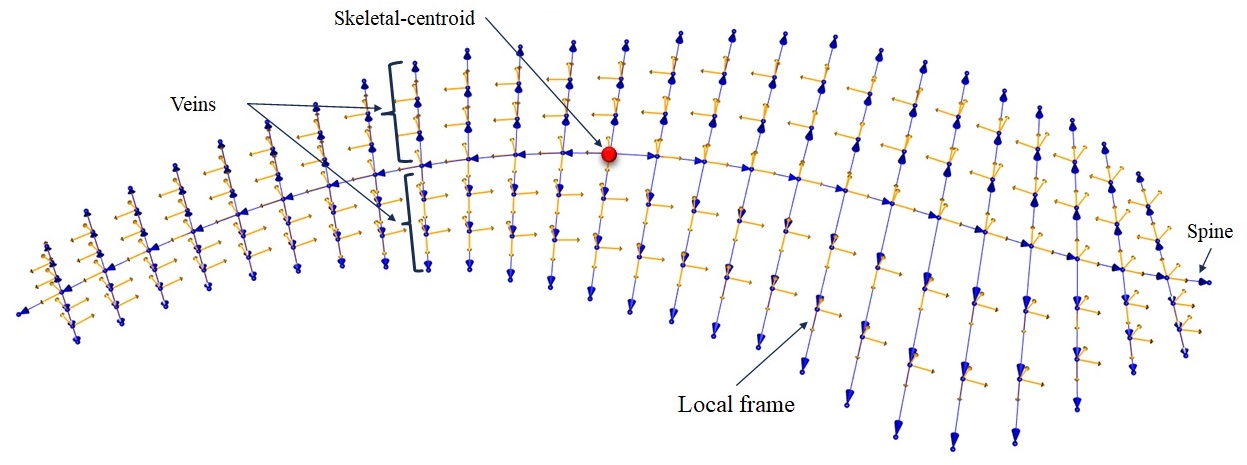}}
\caption{Tree-like structure of the skeletal sheet of a caudate, including local frames in yellow and connection vectors in blue.}
\label{fig:skeletal_sheet}
\end{figure}
Let $\tilde{M}_{\Omega_3}$ be the skeletal sheet of the regular 3EO ${\Omega_3}$. For each point $\bm{p}$ inside $\tilde{M}_{\Omega_3}$, there is a line segment based on the quasi-chordal structure of the cross-sections that contains $\bm{p}$ with endpoints at $\partial\Omega_3^\pm$. We consider one up and one down spoke along the line segment with a tail at $\bm{p}$ and tips on ${\partial\Omega_3}^+$ and ${\partial\Omega_3}^-$, respectively. Thus, the length of the up and down spokes reflects the local width of ${\Omega_3}$, and their difference reflects the local symmetry relative to $\tilde{M}_{\Omega_3}$ (i.e., higher differences indicate higher local asymmetry). Assume the middle point of the spine is the \textit{skeletal-centroid}, as depicted in \Cref{fig:skeletal_sheet}. Thus, the spine can be seen as two curves with the starting point at the skeletal-centroid. Also, the center curve of each cross-section consists of two typically non-straight curves called veins emanating from the spine. Thus, we can consider a vein's starting point on the spine and its ending point on the 3EO's crest. Since $\bm{p}$ is on the skeletal sheet, it belongs to a curve $\gamma\subset\tilde{M}_{\Omega_3}$, where $\gamma$ is the spine or a vein.\par
We define the orthogonal local frame at $\bm{p}$ as $F_{\bm{p}}^\gamma=(\bm{n}_{\bm{p}}^\gamma,\bm{b}^\gamma_{\bm{p}},\bm{b}_{\bm{p}}^{\gamma\perp})\in{SO(3)}$, where $\bm{n}_{\bm{p}}$ is normal to $\tilde{M}_{\Omega_3}$, $\bm{b}^\gamma_{\bm{p}}$ is the velocity vector tangent to $\gamma$, and $\bm{b}_{\bm{p}}^{\gamma\perp}=\bm{n}_{\bm{p}}^\gamma\times\bm{b}_{\bm{p}}^\gamma$. For points on the intersections of the spines and cross-sections' center curves (i.e., veins' starting points), we consider $\gamma$ as the spine. We do not consider frames at the endpoints of the spine and veins, as they are at the edge of the skeletal sheet. Each frame has only one parent frame but up to three child frames. A vector connecting a frame to its parent is called a \textit{connection vector}, with the tail at the origin of the parent frame and the tip at the origin of the (child) frame. In this sense, the union of the connection vectors has a tree-like structure along the spine from the skeletal-centroid out and along the veins from the spine out. This local-frame construction makes the shape representation alignment-independent and enables consistent comparison and statistical analysis across a population without relying on a particular pose or coordinate system, as discussed by \citet{Taheri2022}. \Cref{fig:skeletal_sheet} illustrates the tree-like structure of a caudate's skeletal sheet.\par
We define an LPDSSRep as a tuple of \textit{Geometric Object Properties} (GOPs) of the object as
\begin{equation}\label{equ:LP-ds-rep}
    s=(F^{\ast}_1,...,F^{\ast}_{n_f},\bm{v}^{\ast}_1,...,\bm{v}^{\ast}_{n_c},\bm{u}^{\pm\ast}_1,...,\bm{u}^{\pm\ast}_{n_f},v_1,...,v_{n_c},r_1^\pm,...,r^\pm_{n_f}),
\end{equation}
or in the short form $s=(F^{\ast}_i,\bm{v}^{\ast}_j,\bm{u}^{\pm\ast}_i,v_j,r_i^\pm)_{i,j}$, where $i=1,...,n_f$ and ${j=1,...,n_c}$, $n_f$ is the number of frames, $n_c$ is the number of connection vectors, $F^{\ast}_i\in{SO(3)}$ is the $i$th frame (orientation) based on its parent coordinate system, $\bm{v}^{\ast}_j\in\mathbb{S}^2$, and $v_j\in{\mathbb{R}^+}$ are $j$th connection vector's direction and length, where the direction is based on the local frame (i.e., a frame that tail of the vector is located on), $\bm{u}^{\pm\ast}_i\in\mathbb{S}^2$ are the directions of the $i$th up and down spokes based on their local frames, $r_i^\pm\in{\mathbb{R}^+}$ are the lengths of the up and down spokes at the $i$th frame. Thus, an LPDSSRep lives in a Cartesian product of Euclidean symmetric spaces as 
\begin{equation}\label{equ:LP-ds-rep_space}
    \mathbf{S}=(SO(3))^{n_f}\times(\mathbb{S}^2)^{n_c+2n_f}\times(\mathbb{R}^+)^{n_c+2n_f}.
\end{equation}

\begin{figure}[ht]
\centering
\boxed{\includegraphics[width=0.97\textwidth]{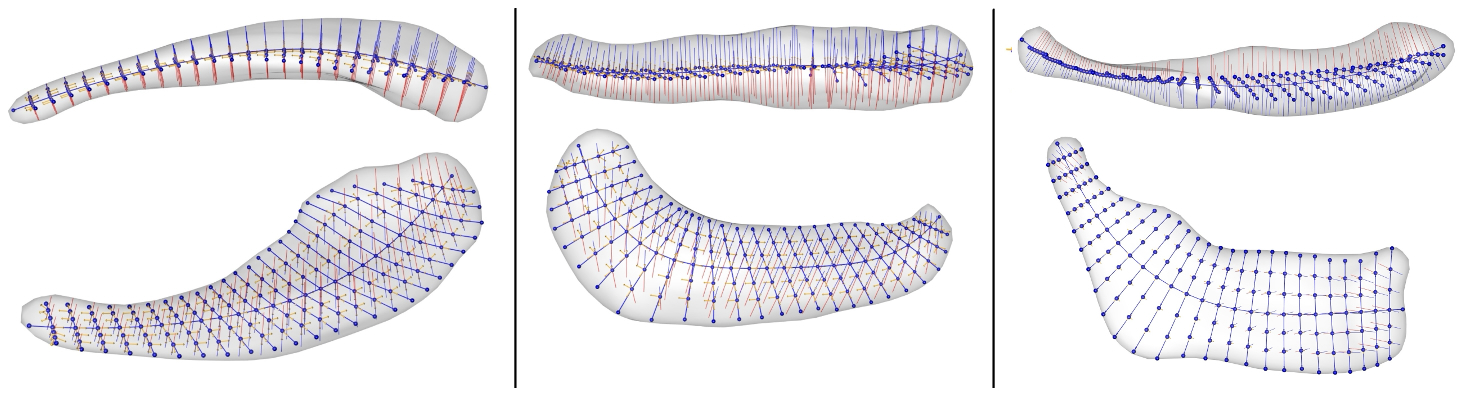}}
\caption{Fitted LPDSSRep to a caudate (left), a hippocampus (middle), and a hemimandible (right), including local frames and spokes.}
\label{fig:caudateHippo3DSpokes}
\end{figure}

Based on our model fitting $\bm{u}^{+\ast}_i=-\bm{u}^{-\ast}_i$, we can ignore down spokes' directions. Thus, we rewrite $s=(F^{\ast}_i,\bm{v}^{\ast}_j,\bm{u}^{+\ast}_i,v_j,r_i^\pm)_{i,j}$. Further, if the skeletal sheet of a regular 3EO has low curvature everywhere, we can consider up and down spokes normal to the skeletal sheet. Therefore, the first frame element at each point represents the direction of its up and down spokes, and we have $s=(F^{\ast}_i,\bm{v}^{\ast}_j,v_j,r_i^\pm)_{i,j}$ and consequently $\mathbf{S}=(SO(3))^{n_f}\times(\mathbb{S}^2)^{n_c}\times(\mathbb{R}^+)^{n_c+2n_f}$. However, to avoid ambiguity, we stick to the definition of \Cref{equ:LP-ds-rep} and \Cref{equ:LP-ds-rep_space} as they cover both the LPDSRep and the LPDSSRep. \Cref{fig:caudateHippo3DSpokes} depicts the fitted LPDSSRep to a caudate (left), a hippocampus (middle), and a hemimandible (right).

The LPDSSRep can be used for both shape analysis (i.e., after removing the size) and size-and-shape analysis (i.e., by preserving the size) \citep{dryden2016statistical}. Since the LPDSSRep is invariant to rigid transformation, for shape analysis, we remove the scale by dividing the vectors' lengths by the size of the LPDSSRep, which we call \textit{LP-size} \citep{Taheri2022}. The LP-size is the geometric mean of the vectors' lengths as
\begin{equation*}
   \|s\|_{LP}=\exp{(\frac{1}{2n_f+n_c}(\sum_{j=1}^{n_c}\ln(v_j)+\sum_{i=1}^{n_f}\ln(r^+_i)+\sum_{i=1}^{n_f}\ln(r^-_i)))}.
\end{equation*}
In \Cref{sec:Evaluation}, we analyze real data based on LPDSSRep shape analysis according to features given earlier.

\section{Goodness of fit}\label{sec:Goodness_of_fit}
In \Cref{sec:skeletal_structure}, we discussed the 3EO's relaxed skeletal structure by fitting the skeletal sheet close to the CVS and defining the spine by fitting a smooth curve close to the relaxed CVS of the flattened skeletal sheet. Thus, we need to explain the level of relaxation. Also, we explored the possibility of choosing different spines based on the chordal skeleton or the relaxed CVS of the flattened sheet (see \Cref{fig:caudateHippoLocalFrames}). In addition, we know by definition that the swept skeletal structure of a 3EO is not unique. Therefore, we need to discuss which model is superior for establishing correspondence and statistical analysis. Obviously, we prefer a model that represents the object such that the implied boundary of the skeletal model approximates the actual boundary. Also, the model should be locally as symmetric as possible because symmetry is the key element of any skeletal structure. Further, it should be tidy to avoid false positives in the analysis, as we discuss in \Cref{sec:Evaluation}.\par
Thus, it is reasonable to search for a superior model with a tidier and more symmetric structure. This can be done by tuning the model fitting procedure based on the flexibility of the curve or surface fitting methods (e.g., by modifying the number of basis functions in B-spline or Fourier expansion \citep{kokoszka2017introduction} or by changing the degree of the polynomial in \textit{polynomial regression} \citep{jorgensen1993theory,montgomery2015introduction}). In this section, we discuss the goodness of fit of an LPDSSRep by considering its \textit{volume-coverage}, \textit{skeletal-symmetry}, and \textit{skeletal-tidiness}. Then, in \Cref{sec:Example}, we discuss our fitting strategy by an example.

\subsection{Volume-coverage}\label{sec:volume_Coverage}
Let $s$ be a fitted LPDSSRep to 3EO $\Omega$, where its spokes' tips are located on $\partial\Omega^\pm$, and the endpoints of its veins are located on $\partial\Omega^0$. Therefore, we have a point cloud of the spokes' tips and veins' endpoints representing the implied boundary of $s$. We use this point cloud to generate the implied boundary as a triangular mesh. Specifically, we interpolate the spokes within the quadrilateral patches of the skeletal sheet to generate the implied boundary, as discussed by \citet{han2006interpolation,liu2021fitting}.\par
Let $\hat\Omega$ be the object generated by the implied boundary. One important factor for the goodness of fit is to see how well $\hat\Omega$ approximates $\Omega$. In other words, we prefer a model in which the implied boundary is as close as possible to the actual boundary. Thus, we consider the volume-coverage as the Jaccard index of the model, as $J(\Omega,\hat\Omega)\in[0,1]$, where $J(\Omega,\hat\Omega)=1$ reflects identical volume. Obviously, we prefer models with higher volume-coverage (see \Cref{fig:goodnessOfFit}).

\subsection{Skeletal-symmetry}
According to the definition of 3EOs, the spine is located on the center curve of each cross-section, approximately at its middle point. The union of the cross-sections' center curves forms the object's skeletal sheet. Thus, the skeletal sheet represents the 3EO's skeleton, and the spine can be seen as a nonlinear skeleton of the skeletal sheet \citep{Taheri2022}.\par
In LPDSSRep, for each up spoke $\bm{s}_{(\bm{p},\bm{u}^+)}$ we have a down spoke $\bm{s}_{(\bm{p},\bm{u}^-)}$ with lengths $r^+=\|\bm{s}_{(\bm{p},\bm{u}^+)}\|$ and $r^-=\|\bm{s}_{(\bm{p},\bm{u}^-)}\|$, respectively. Also, the spine divides the skeletal sheet into two parts, namely, the \textit{right side} and the \textit{left side}. For each vein $\nu_{\bm{p}_s}^+$ on the right side, we have a coplanar vein on the left side $\nu_{\bm{p}_s}^-$ with a common starting point $\bm{p}_{sp}$ on the spine. Thus, in a symmetric model, we expect $r^+\sim{r^-}$, and $\|\nu_{\bm{p}_{sp}}^+\|_g\sim\|\nu_{\bm{p}_{sp}}^-\|_g$, where $\|.\|_g$ measures the curve length. Assume we have $n_v$ veins on each side of the spine. Assume vector $\bm{l}^+=(r_1^+,...,r_{n_f}^+,\|\nu_1^+\|_g,...,\|\nu_{n_v}^+\|_g)^T$, where $\nu_t^+$ is the $t$th vein on the right side and $t\in\{1,...,n_v\}$. Similarly, assume $\bm{l}^-=(r_1^-,...,,r_{n_f}^-,\|\nu_1^-\|_g,...,\|\nu_{n_v}^-\|_g)^T$. For simplicity, we write $\bm{l}^+=(l^+_1,...,l^+_{n_f+n_v})^T$ and $\bm{l}^-=(l^-_1,...,l^-_{n_f+n_v})^T$. We define \textit{skeletal length} as $l_s=\sum_{i=1}^{n_f+n_v}l^+_i+\sum_{i=1}^{n_f+n_v}l^-_i$ and define the weight vector $\bm{w}=(w_1,...,w_{n_f+n_v})^T$ such that $w_i=\frac{l^+_i+l^-_i}{l_s}$. Then, we define skeletal-symmetry as $\sum_{i=1}^{n_f+n_v}w_i\cdot{f_i}$, where $f_i=\frac{l^+_i}{l^-_i}$ if $l^-_i\geq{l^+_i}$ and otherwise ${f_i=\frac{l^-_i}{l^+_i}}$. Thus, skeletal-symmetry is $\in[0,1]$. If we have a perfect symmetry (i.e., ${\forall{i}: l^+_i={l^-_i}}$), then skeletal-symmetry is 1.0.

\subsection{Skeletal-tidiness}\label{sec:SkeletonPerturbation}
A model with good volume-coverage and skeletal-symmetry may still be messy. Obviously, we prefer skeletal models that are as tidy as possible for two main reasons. First, untidy models increase the number of false positives and make the analysis misleading because the analysis is a frame-by-frame analysis. Since we consider frames on the skeletal sheet, even a small perturbation on the skeletal sheet violates frame correspondence and hugely affects the results, as we demonstrate by a simple example in \Cref{sec:ellipsoid_simulation}. Thus, we prefer a model such that the spine and cross-sections' center curves have the lowest possible curvature. Second, as discussed in \Cref{sec:introduction}, the center curve at each point must satisfy the RCC.\par
Let us assume a moving frame $F^\gamma$ on an open curve $\gamma$ (representing the spine or a vein) located on the skeletal sheet $M$. In discrete format we assume a sequence of $N+1$ equidistant points $\bm{p}_0,...,\bm{p}_N$ on $\gamma$ and define \textit{average perturbation} of $\gamma$ as $\zeta(\gamma)=\frac{1}{N}\sum_{i=1}^{N}d_g(\bm{q}_{i-1},\bm{q}_{i})\in[0,\frac{\pi}{2}]$, where $\bm{q}_{i}\in\mathbb{S}^3$ is the unit quaternion representation of the frame $F_{\bm{p}_i}^\gamma$, and $d_g(\bm{x},\bm{y})=\arccos(\abs{\bm{x}^T\bm{y}})$ is the first quadrant geodesic distance \citep{huynh2009metrics}. Thus, $\zeta(\gamma)$ is the mean integrated rotation of the moving frame along $\gamma$ when $N\rightarrow{\infty}$.\par
Assume a discrete skeletal sheet $M$ with a spine $\gamma_{s}$, spinal points $\bm{p}_0,...,\bm{p}_{N+1}$, and $N$ cross-sections' center curves $\{\gamma_{i}\}_{i=1}^{N}$ such that $\gamma_{s}\cap\gamma_{i}=\bm{p}_i$. We know $\forall{i}:$ $\bm{n}_{\bm{p}_i}^{\gamma_s}=\bm{n}_{\bm{p}_i}^{\gamma_{i}}$. We define the degree of rotation of the $i$th cross-section relative to the spine by $d_r(\gamma_s,\gamma_{i})=d_g(\bm{b}_{\bm{p}_i}^{\perp\gamma_s},\bm{b}_{\bm{p}_i}^{\gamma_i})$. Finally, the average perturbation of the skeletal sheet $M$ is defined as
\begin{equation*}
\zeta^\dagger(M)=\frac{2}{(2N+1)\pi}\left(\zeta({\gamma_{s}})+\sum_{i=1}^N\zeta({\gamma_{i}})+\sum_{i=1}^Nd_r(\gamma_s,\gamma_{i})\right)\in[0,1),   
\end{equation*}
and the average tidiness of $M$ as $1-\zeta^\dagger(M)$. Thus, the average tidiness of the ellipsoid's skeleton is 1.0 if the spine is the ellipsoid's major axis, and the cross-sections (with straight center curves) are normal to the spine, i.e., $\forall{i}:d_r(\gamma_s,\gamma_i)=0$.\par
In practice, we need the skeletal sheet as tidy as possible. Thus, it is reasonable to consider $\zeta(\gamma)=\frac{2}{\pi}\max\{d_g(\bm{q}_{i-1},\bm{q}_{i})\}_{i=1}^N$ and consequently define \textit{strict tidiness} as $1-\zeta^\ddagger(M)$ to judge a model based on its most disordered element, where $\zeta^\ddagger(M)=\frac{2}{\pi}\max\{\zeta(\gamma_s),\max\{\zeta(\gamma_i)\}_{i=1}^N,\max\{d_r(\gamma_s,\gamma_i)\}_{i=1}^N\}$. We define the goodness of fit score as the multiplication of volume-coverage, skeletal-symmetry, and (average or strict) tidiness.  In this sense, a model is superior if its score is closer to 1.0.

\subsection{Example}\label{sec:Example}
The flexibility of our method comes from fitting the skeletal sheet $\tilde{M}_{\Omega_3}$ and the center curve of the flattened skeletal sheet $\Gamma'_2$. 
Although it is possible to define a standard approach to make the fitting procedure straightforward by, for example, fitting $\tilde{M}_{\Omega_3}$ using the B-spline surface fitting of \citet{lee1997scattered} and fitting $\Gamma'_2$ using the principal curve fitting of \citet{hastie1989principal}, such a standard approach might not be flexible enough to generate distinct LPDSSReps for a 3EO. Therefore, in this work, we prefer to use polynomial regression (PR). Thus, by changing the PR degree from 1 to $n$ (where we consider $n\leq{7}$), we can fit $n\times{n}$ different LPDSSReps to a 3EO. Then, we select the best model based on the goodness of fit. We leave a detailed discussion about choosing the best curve or surface fitting method (from various available methods such as implicit polynomial \citep{unsalan1999new}, nonlinear regression \citep{ritz2008nonlinear}, etc.) for our future studies.
\begin{figure}[ht]
\centering
\boxed{\includegraphics[width=0.98\textwidth]{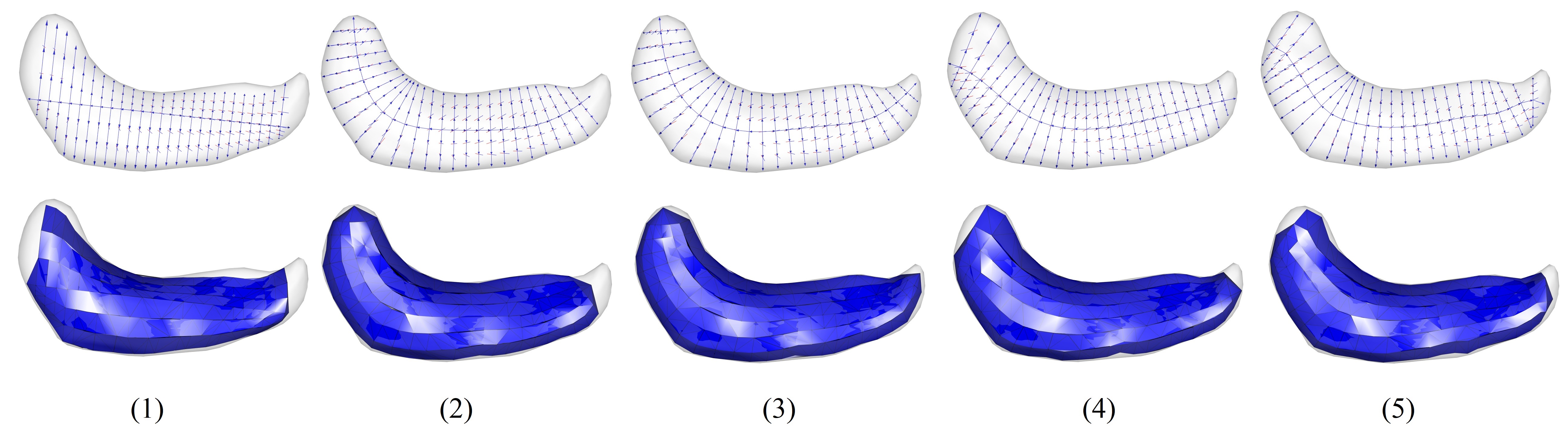}}
\caption{Top row: Fitted LPDSSReps with parameters given in \Cref{Table_GoodnessOfFit}. Bottom: Implied boundary in blue.}
\label{fig:goodnessOfFit}
\end{figure}
\begin{table}[ht]
\caption{Goodness of fit.}
\centering
\resizebox{\textwidth}{!}{%
\begin{tabular}{@{}cccccccc@{}}
\toprule
Fit & \begin{tabular}[c]{@{}c@{}}PR degrees \\ $\tilde{M}_{\Omega_3}$, $\Gamma'_2$\end{tabular} & Volume-coverage & skeletal-symmetry & Average tidiness $\zeta^\dagger$ & Strict tidiness $\zeta^\ddagger$ & Score 1 & Score 2\\ \midrule

\#1 & 1, 1 & 0.812 & 0.724 & 1     & 1     & 0.588 & 0.588 \\ \midrule
\#2 & 3, 4 & 0.877 & 0.883 & 0.969 & 0.861 & 0.750 & 0.667 \\ \midrule
\#3 & 4, 4 & 0.890  & 0.873 & 0.973 & 0.890  & 0.756 & 0.692 \\ \midrule
\#4 & 4, 5  & 0.876 & 0.861 & 0.972 & 0.870  & 0.733 & 0.656 \\ \midrule
\#5 & 7, 7 & 0.878 & 0.894 & 0.971 & 0.735 & 0.762 & 0.577 \\ \bottomrule

\end{tabular}
}
\label{Table_GoodnessOfFit}
\end{table}
\Cref{fig:goodnessOfFit} depicts 5 (from 49) fitted LPDSSReps to a hippocampus mesh in the top row and their corresponding implied boundary in the bottom row in blue. The volume of the hippocampus mesh is 3425.4 $\mathrm{mm}^3$. The PR degrees for fitting the $\tilde{M}_{\Omega_3}$ and $\Gamma'_2$ are shown in \Cref{Table_GoodnessOfFit}. Note that the boundary division and, consequently, the CVS for all five samples are the same. For the fit \#1 we have a straight line segment as the spine on a perfectly flat skeletal sheet. Although the model is perfectly tidy, its volume-coverage and skeletal-symmetry are weak, resulting in a low model score; see \Cref{Table_GoodnessOfFit}, where score 1 is based on average tidiness and score 2 is based on strict tidiness.\par
By adding more flexibility to the spine and the skeletal sheet, we obtained models \#2-5. The best skeletal-symmetry and score 1 value belong to model \#5. However, as is obvious from \Cref{fig:goodnessOfFit} (5), we have a few sudden changes in spinal frames (i.e., cross-sections with high rotation degrees), which reduce the strict tidiness. In contrast, based on score 2, model \#3 is the winner as it has good volume-coverage ($\approx90\%$), skeletal-symmetry ($\approx87\%$), and strict tidiness ($\approx90\%$) as we do not observe any extreme changes in frame rotations.

\section{Hypothesis testing}\label{sec:Statistical_analysis}
One of the main objectives of LPDSSRep analysis is to detect dissimilarities between two groups of objects. We explain hypothesis testing for LPDSSReps similar to LPDSReps \citep{Taheri2022}. To do so, in this section, we discuss global and partial LPDSSRep hypothesis testing based on LPDSSRep Euclideanization. In addition to the hypothesis testing, LPDSSRep classification is explained in \supplementary.\par
Let $\bm{q}^\ast_i\in\mathbb{S}^3$ be the unit quaternion representation of the $i$th frame \citep{huynh2009metrics}. From \Cref{equ:LP-ds-rep} and \Cref{equ:LP-ds-rep_space} we have $s=(\bm{q}^{\ast}_i,\bm{v}^{\ast}_j,\bm{u}^{\pm\ast}_i,v_j,r_i^\pm)_{i,j}$ and $\mathbf{S}=(\mathbb{S}^3)^{n_f}\times(\mathbb{S}^2)^{n_c+2n_f}\times(\mathbb{R}^+)^{n_c+2n_f}$, where $\mathbb{S}^3$ is the space of local frames based on their unit quaternion representations. Therefore, for a population of LPDSSReps, we have $n_c+3n_f$ sets of spherical data (on $\mathbb{S}^2$ and $\mathbb{S}^3$) and $n_c+2n_f$ sets of positive real numbers. The spherical data can be Euclideanized by \textit{Principal Nested Spheres} (PNS) \citep{jung2012analysis} or to reduce the computational cost, by projecting the data to the tangent space (which we use in this work), as discussed by \citep{kim2020kurtosis}. Given the mapping $\mathscr{F}:\mathbb{S}^d\rightarrow{\mathbb{R}^d}$, $s^{e}=(\mathscr{F}(\bm{q}^{\ast}_i),\mathscr{F}(\bm{v}^{\ast}_j),\mathscr{F}(\bm{u}^{\pm\ast}_i),\log{v_j},\log{r_i^\pm})_{i,j}$ is the Euclideanized version of $s$ from \Cref{equ:LP-ds-rep} that lives on the product space $(\mathbb{R}^3)^{n_f}\times(\mathbb{R}^2)^{n_c+2n_f}\times(\mathbb{R}^+)^{n_c+2n_f}$.\par

Now, assume $A=\{{s^{e}_A}_m\}_{m=1}^{N_1}$ and $B=\{{s^{e}_B}_m\}_{m=1}^{N_2}$ are two groups of Euclideanized LPDSSReps (or LPDSRep) of sizes $N_1$ and $N_2$. We can consider the vectorized version of LPDSSReps in the feature space (see classification in \supplementary ) and design a global test to compare $A$ and $B$ as two multivariate distributions. Or we can consider LPDSSReps as tuples and compare the two populations of tuples element-wise to detect locational dissimilarities, as discussed by \citet{Taheri2022}.

\subsection{Global test}\label{sec:global_test}
Let $\mu_A$ and $\mu_B$ be the means of the sets $A$ and $B$, respectively. For the global test, we test $H_0:\mu_A=\mu_B$ versus $H_0:\mu_A\neq\mu_B$. Since the feature space is a high-dimensional space, we apply \textit{Direction Projection Permutation} (DiProPerm) \citep{wei2016direction} with \textit{Distance Weighted Discriminator} (DWD) \citep{marron2007distance}.

\subsection{Partial test}\label{sec:patial_test}
The main objective of LPDSSRep is to detect local dissimilarities. Thus, we design partial hypothesis tests on the GOPs (of \Cref{equ:LP-ds-rep}). Basically, when the result of the global test shows significant dissimilarities, we apply partial tests to detect and explain local dissimilarities.\par
Let $n_{t}$ be the total number of GOPs. To test GOPs' mean difference, we design $n_{t}$ partial tests. Let $\mu_{A}(k)$ and $\mu_{B}(k)$ be the observed sample mean of the $k$th GOP from $A$ and $B$, respectively. The partial test is ${H_{0k}:\mu_{A}(k)=\mu_{B}(k)}$ versus ${H_{1k}:\mu_{A}(k)\neq\mu_{B}(k)}$, where $k\in\{1,...,n_t\}$. For the partial test, we can apply Hotelling's T$^2$ test with the normality assumption or a permutation test. Finally, we need to control false positives by methods such as the \textit{family-wise error rate} of \citet{bonferroni1936teoria} or the \textit{False Discovery Rate} (FDR) of \citet{benjamini1995controlling} (BH).

\section{Evaluation}\label{sec:Evaluation}
The proposed LPDSSRep is evaluated in comparison to the available LPDSRep, which is based on the fitting method of \citet{liu2021fitting} based on visual inspection and statistical analysis.

\subsection{Visual inspection}
\begin{figure}[ht]
\centering
\boxed{\includegraphics[width=0.90\textwidth]{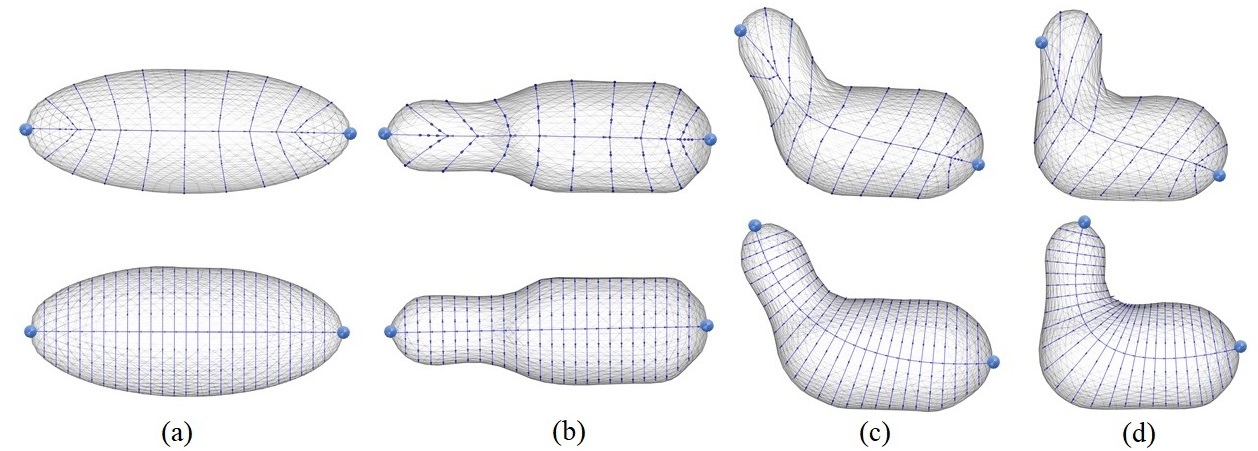}}
\caption{Toy example illustrating differences in the fitting behavior of the LPDSRep (top row) and LPDSSRep (bottom row) for increasingly bent 3EOs. Bold blue points indicate the spine endpoints.}
\label{fig:deformationIssue}
\end{figure}
For the visual inspection, we first use a toy example to illustrate how the different fitting procedures affect the resulting skeletal structures.
To design the toy examples, we slightly deform an ellipsoid to make a 3EO as a reference object, as illustrated in column (b) of \Cref{fig:deformationIssue}. The reference object can be seen as an arm. By bending the reference object with different degrees of bending at the elbow, we make two other 3EOs, as depicted in the figure's columns (c) and (d). To obtain LPDSReps, we used the SlicerSALT toolkit \citep{vicory2018slicersalt}. As the bending increases, the LPDSRep (top row) does not adequately follow the deformation of the reference object, as its spine does not bend sufficiently to capture the increasing curvature. Moreover, the skeletal sheet becomes increasingly irregular. A possible explanation could be the poor boundary registration or the effect of boundary deformation on the skeletal sheet. In contrast to the LPDSRep, we can see that the structure of the skeletal sheet in the LPDSSRep is tidier as depicted in the bottom row of \Cref{fig:deformationIssue}. Also, the spine shows more flexibility, such that the endpoints of the spines (depicted by bold blue points) are located at the points corresponding to the vertices of the ellipsoid.\par
\begin{figure}[ht]
\centering
\boxed{\includegraphics[width=0.97\textwidth]{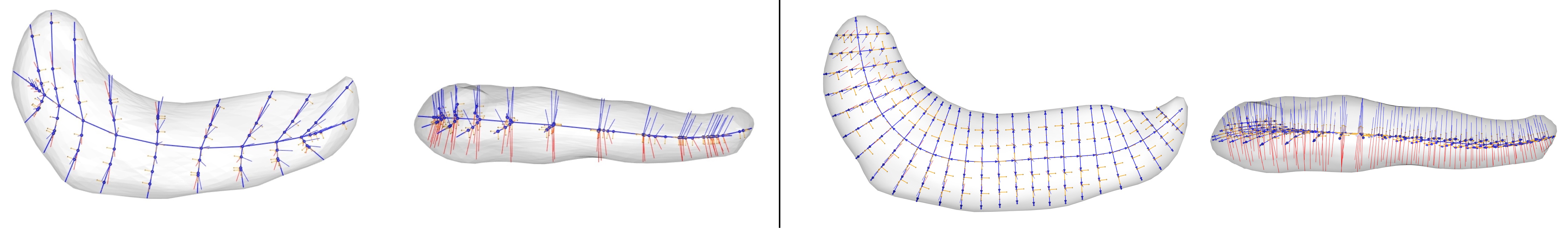}}
\caption{Fitted LPDSRep (left) and LPDSSRep (right) in a hippocampus in two angles.}
\label{fig:hippoCaudDeformationIssue}
\end{figure}
Analogous to the toy examples, we can see the same issues of the LPDSRep on real data. \Cref{fig:hippoCaudDeformationIssue} illustrates the LPDSReps of hippocampus in two angles with the LPDSRep (left), and the LPDSSRep (right). It seems the spine provides a better description of the 3EO center curve in LPDSSRep. 

\subsection{Statistical analysis}
For the statistical analysis, we start with a simulation of a toy example before analyzing real data.
\subsubsection{Toy example}\label{sec:ellipsoid_simulation}
\begin{figure}[ht]
\centering
\boxed{\includegraphics[width=0.97\textwidth]{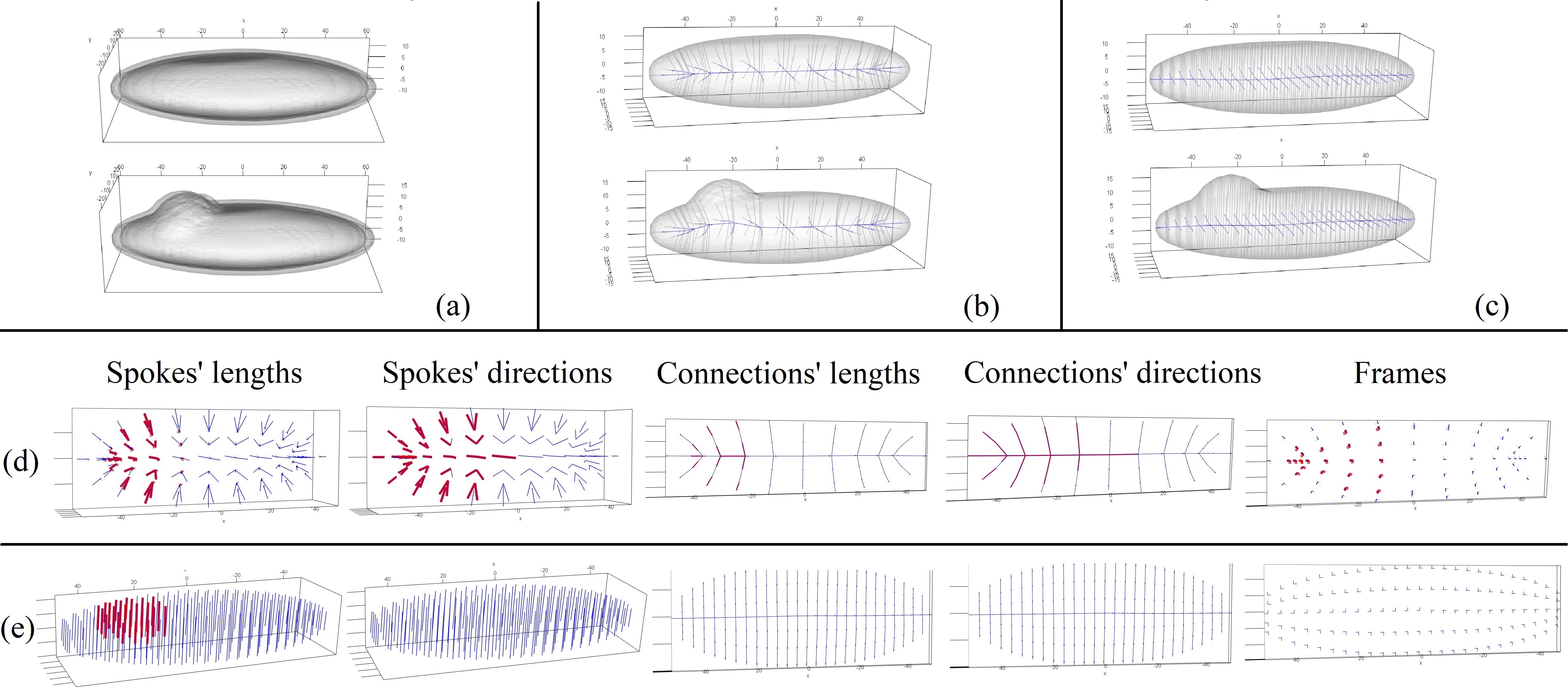}}
\caption{(a) Two groups of simulated ellipsoids with and without boundary protrusion. (b) and (c) LPDSRep and LPDSSRep of two samples from each group. (d) and (e) results of the partial tests on LPDSRep and LPDSSRep, respectively. Red indicates significant GOPs.}
\label{fig:simulationEllipsoid}
\end{figure}

We design a very simple example by simulating two groups of fifty ellipsoids based on randomly generated principal radii such that the ellipsoids of the first group are perfectly symmetric. In contrast, in the second group, there is a small protrusion on the boundary, as illustrated in \Cref{fig:simulationEllipsoid} (a). \Cref{fig:simulationEllipsoid} (d) and (e) show the results of the partial tests from \Cref{sec:patial_test} on the fitted LPDSReps and LPDSSReps, respectively. Although both methods point to the area where we expect to see the difference, the LPDSSRep reflects the dissimilarity more accurately based on the length of the top spokes in the critical region. In fact, the LPDSRep introduces more significant GOPs such that we may conclude that the left part of the two groups is totally different. The reason is that there is a perturbation of the skeletal sheet caused by the boundary deformation, as depicted in \Cref{fig:simulationEllipsoid} (b). 
In comparison, LPDSSReps show more resistance against the protrusion and thus better correspondence among the skeletal sheets of the two groups
(see \Cref{fig:simulationEllipsoid} (c)). The reason is that the CVS reflects the interior structure of the object and, after relaxation, is less affected by local boundary protrusions than a fitting procedure driven directly by boundary deformation.

\subsubsection{Real data}\label{sec:Real_Data}

For the statistical analysis, we compare the left hippocampus of 182 patients with early Parkinson's disease (PD) with the left hippocampus of 108 healthy people as a control group (CG). 
The data is provided by the Stavanger University Hospital (\url{https://helse-stavanger.no}) and the ParkWest study (\url{http://parkvest.no}).
For this, LPDSSReps are fitted to the hippocampi of both populations and compared with global and partial hypothesis tests as described in \Cref{sec:global_test} and \Cref{sec:patial_test} to evaluate if there are differences between the two populations.\par
\begin{figure}[ht]
\centering
\boxed{\includegraphics[width=0.98\textwidth]{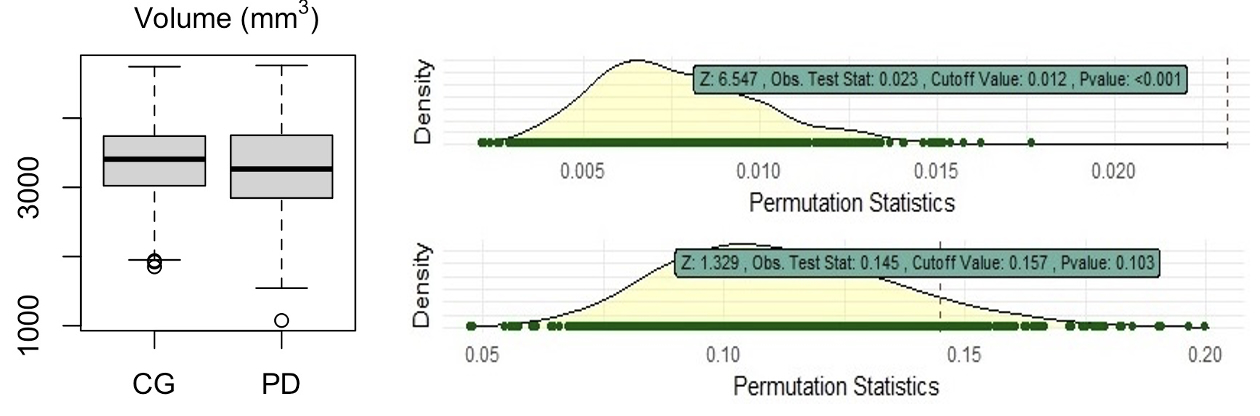}}
\caption{Left: Volume comparison box plot. Right: DiProPerm plots of LPDSSRep (top) and LPDSRep (bottom).}
\label{fig:DiProPerm}
\end{figure}

The volumetric analysis showed substantial overlap in hippocampal volumes between the CG and PD groups, as illustrated by the box plot in \Cref{fig:DiProPerm} (left). Although the CG group exhibits a slightly higher mean volume, the t-test showed that this difference was not statistically significant, as the $p$-value is 0.258. The mean hippocampal volumes are 3352.2 $\mathrm{mm}^3$ for the CG and 3271.4 $\mathrm{mm}^3$ for the PD.

\paragraph{Global test}

\Cref{fig:DiProPerm} (right) presents the results of the global tests based on the DiProPerm analysis comparing LPDSRep and LPDSSRep. The $p$-value is less than $0.001$ for LPDSSReps and $0.103$ for LPDSReps. Thus, given a significance level of $\alpha=0.05$, PD and CG are statistically significantly different based on LPDSSReps. Notably, LPDSSReps have better performance compared to LPDSReps based on the higher Z-score (6.54 vs. 1.32), which indicates that the data are more separated for LPDSSReps. This might be the effect of the above-described stiffness of LPDSReps, their sensitivity to local boundary protrusions, or a combination of both. Consistent with the overall comparison, LPDSReps also show poorer classification performance, as discussed in the \supplementary.

\begin{figure}[ht]
\centering
\boxed{\includegraphics[width=0.97\textwidth]{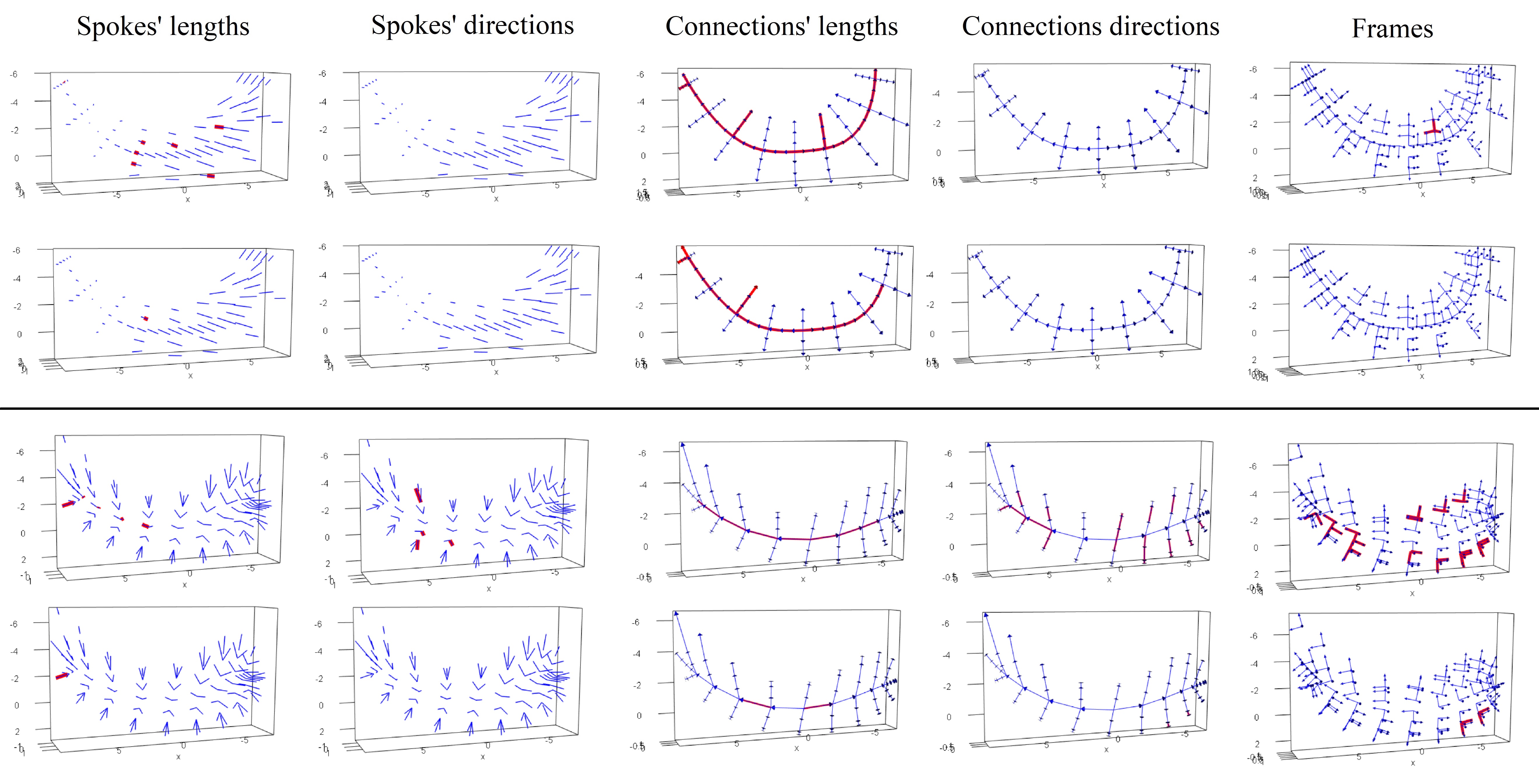}}
\caption{Partial tests. Red indicates significant GOPs. Results based on the LPDSSRep (Top) and the LPDSRep (Bottom) before and after BH $p$-value adjustment in the first and second rows, respectively. Significant level $\alpha=0.05$ and FDR=0.1.}
\label{fig:LocalTests_RealData}
\end{figure}

\paragraph{Partial tests}
The partial tests are illustrated in \Cref{fig:LocalTests_RealData}. Both the LPDSRep and the LPDSSRep show significant differences in the spinal connection vectors. The LPDSSRep introduces fewer significant connection vectors' directions and frames' orientations. This behavior supports our conclusion about the skeletal sheet from the toy example.

\section{Conclusion}\label{sec:Conclusion}
In this work, we introduced a novel Locally Parameterized Discrete Swept Skeletal Representation (LPDSSRep) for 3D Ellipsoidal Objects (3EOs). The proposed representation is designed to provide meaningful correspondences across a population of 3EOs, which is essential for statistical shape analysis. The fitting framework is based on partitioning the 3EO boundary into two components, constructing a skeletal sheet from the relaxed Central Voronoi Skeleton (CVS), and establishing a tree-like parameterization of the skeletal sheet through mappings between $\mathbb{R}^3$ and $\mathbb{R}^2$. To objectively select the most appropriate skeletal model, we introduced a set of goodness-of-fit criteria based on volume coverage, skeletal symmetry, and skeletal tidiness. These criteria enable the evaluation and selection of the candidate model that best represents the target object while preserving its swept region characteristics.\par
The proposed LPDSSRep was evaluated on both synthetic toy examples and real hippocampal datasets. Compared with the existing LPDSRep, the proposed representation demonstrated superior performance in terms of visual correspondence quality and statistical sensitivity in hypothesis testing. 

Although previous studies have compared s-rep-based representations with alternative shape correspondence methods \citep{Tu2016, Tu2018}, an updated direct comparison of the proposed LPDSSRep with current state-of-the-art shape analysis methods would be an interesting direction for future work. From a theoretical perspective, ensuring non-intersection of the slicing planes and skeletal spokes under arbitrary refinement of the discrete representation warrants further investigation. A related theoretical direction is to study the convergence properties of increasingly refined discrete swept skeletal representations. Finally, extending the proposed framework to broader classes of irregular 3EOs opens further directions for developing swept skeletal models for more general object classes.

The results suggest that, when combined with an appropriate model-fitting strategy, the LPDSSRep provides a robust and anatomically meaningful representation capable of capturing subtle local shape variations across a population. We believe that the proposed framework establishes a promising foundation for the statistical analysis of ellipsoidal anatomical structures. 

\section*{Acknowledgments}
This work is funded by the Department of Mathematics and Physics of the University of Stavanger (UiS). We thank Profs. James Damon (late of UNC), J.S. Marron (UNC), and Jan Terje Kvaløy (UiS) for insightful discussions and inspiration for this work. We also thank Prof. Guido Alves (UiS) for providing ParkWest data.

\section*{Supplementary Material}\label{sec:supplementary}

\subsection*{Choosing $\delta$ for the affinity matrix}
\begin{figure}[ht]
\centering
\boxed{\includegraphics[width=0.97\textwidth]{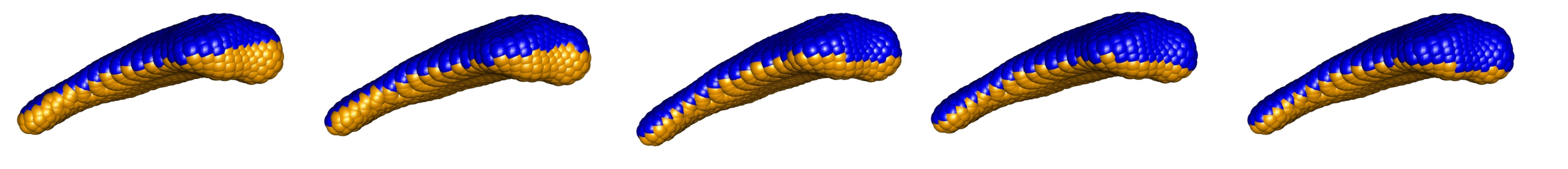}}
\caption{From left to right, the boundary divisions are based on $\delta$ as 0.1, 0.2, 0.5, 0.8, and 0.9, respectively.}
\label{fig:differentDeltas}
\end{figure}

Choosing the value of $\delta$ that defines the affinity matrix can slightly change the boundary division. However, based on our observation for most of the objects of our study, $\delta=0.5$ is a reasonable choice. \Cref{fig:differentDeltas} depicts the boundary division of a caudate based on $\delta$ as 0.1, 0.2, 0.5, 0.8, and 0.9, from left to right, respectively

\subsection*{Normality versus symmetricity}
\begin{figure}[ht]
\centering
\boxed{\includegraphics[width=0.68\textwidth]{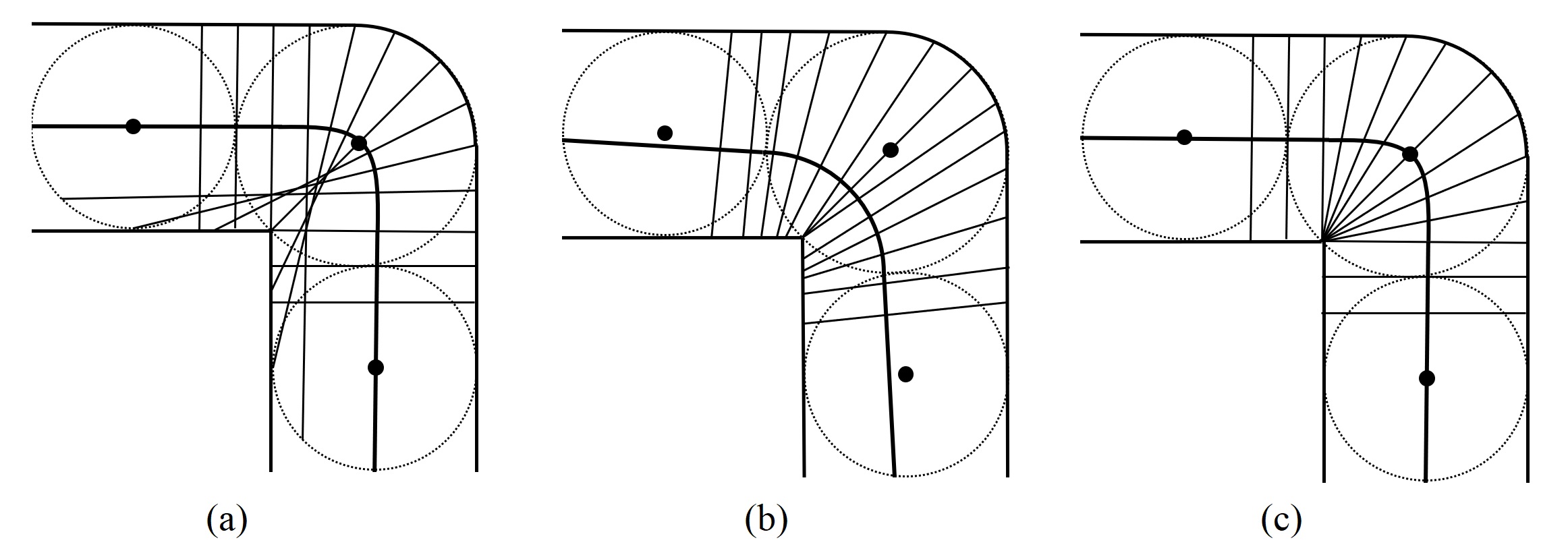}}
\caption{(a) Cross-sections are normal to the medial skeleton. The RCC is violated. (b) Cross-sections are normal to the center curve, but the center curve is not medial. (c) Cross-sections are not normal to the center curve, but the center curve as the chordal axis traverses the cross-section's middle points.}
\label{fig:normality_vs_symmetricity}
\end{figure}
In this section, we use \Cref{fig:normality_vs_symmetricity} (which is similar to Figure 3 of \citep{shani1984splines}) to show even if a 2EO has a non-branching smooth medial skeleton defining a perfectly symmetric model with normality condition is not feasible. In \Cref{fig:normality_vs_symmetricity} (a), the cross-sections are normal to the medial skeleton, but the RCC is violated. In \Cref{fig:normality_vs_symmetricity} (b), cross-sections are normal to the center curve, but the center curve does not meet the middle of the cross-sections. Thus, the model is not symmetric. The model of  \Cref{fig:normality_vs_symmetricity} (c) is perfectly symmetric as the center curve is the chordal skeleton. However, the cross-sections are not normal to the chordal skeleton.

\subsection*{Importance of the RCC}
\begin{figure}[ht]
\centering
\boxed{\includegraphics[width=0.60\textwidth]{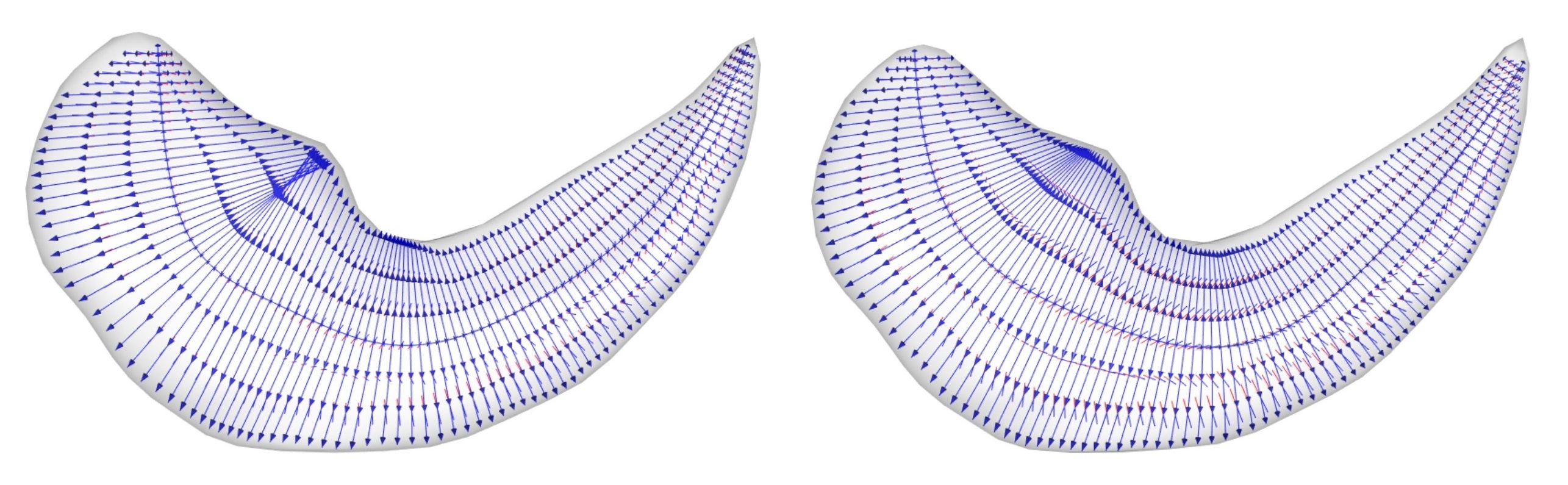}}
\caption{Left: The RCC is violated. A few cross-sections are intersected within the object. Right: The RCC is satisfied.}
\label{fig:curvatureTalorance}
\end{figure}

In this section, we provide an illustration to show the importance of the RCC in defining the spine of a swept skeletal structure. \Cref{fig:curvatureTalorance} depicts two swept skeletal structures of a hippocampus. In \Cref{fig:curvatureTalorance} (right), the RCC is satisfied by the spine while in \Cref{fig:curvatureTalorance} (left) the RCC is violated. While the spines in both figures exhibit notable similarity, it becomes apparent that defining the spine solely as a curve positioned in the middle of the object is insufficient without considering the RCC.

\subsection*{SPHARM-PDM correspondence issue}
\begin{figure}[ht]
\centering
\boxed{\includegraphics[width=0.97\textwidth]{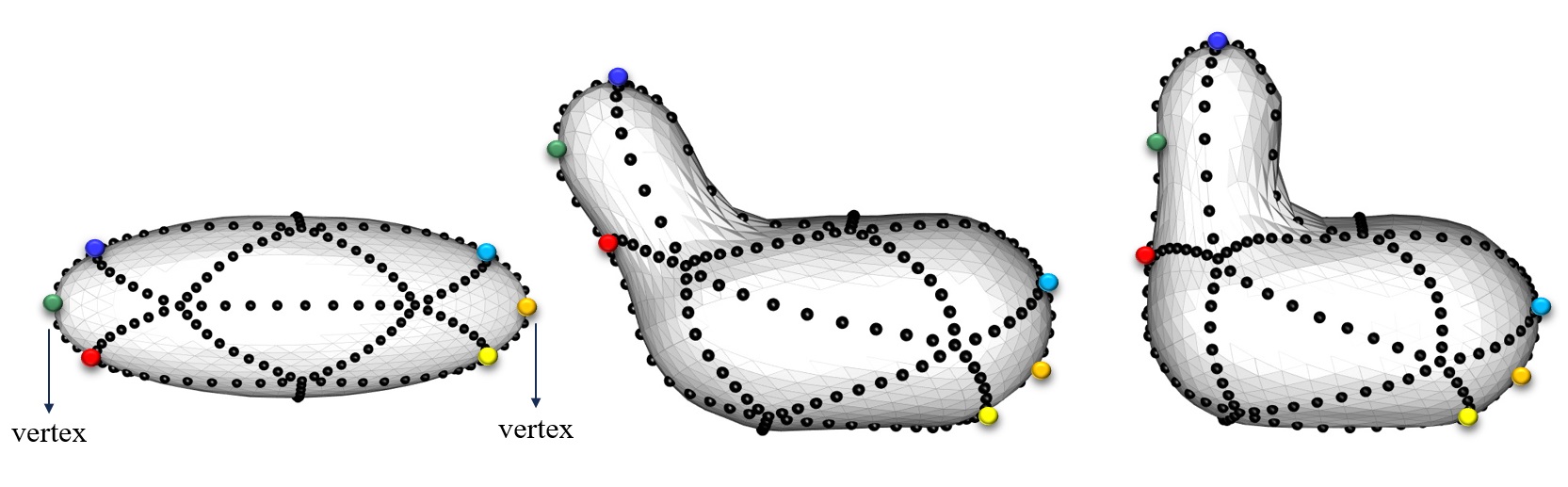}}
\caption{SPHARM-PDM correspondence. Points with the same color are in correspondence.}
\label{fig:SPHARM_PDM}
\end{figure}

As depicted in \Cref{fig:SPHARM_PDM}, SPHARM-PDM fails to define a good correspondence between the boundary of an ellipsoid and the bent versions of it, as the vertices of the ellipsoid do not correspond to the vertices of the objects.

\subsection*{Issues with the curve skeleton}
There are available methods for calculating the spine of 3EOs as a \textit{curve skeleton}. The curve skeleton, as defined by \citet{dey2006defining}, is unique but typically has a branching structure. Based on our experiences, most of the recent methods for pruning and smoothing the curve skeleton, such as Laplacian contraction of \citet{au2008skeleton}, mean curvature skeleton of \citet{tagliasacchi2012mean}, $L1$-medial skeleton of \citet{huang2013l1}, and skeletonization via local separators of \citet{baerentzen2021skeletonization, baerentzen2023multilevel}) are more or less suitable for tube-like objects with circular cross-sections. But, they do not provide a suitable spine for ellipsoidal objects. The primary issue is that almost all of these methods blindly search for the curve skeleton without considering important boundary properties, such as the crest. We believe a proper spine should be on the skeletal sheet, connecting the 3EO's vertices together, and be relatively in the middle of the cross-sections'. Based on our observations, the mentioned methods do not provide such a spine, even for simple 3EOs. In addition, they entirely ignore the RCC. The RCC is not a prerequisite for calculating the curve skeleton because the curve skeleton does not necessarily represent the center curve of a swept region. However, the RCC is vital for defining the center curve of a swept region (see \Cref{fig:curvatureTalorance}). Based on our observations, the curve skeleton exhibits unpredictable behavior inside a 3EO. For example, it bends and swings freely inside the object. Therefore, defining corresponding skeletal structures based on the curve skeleton is highly questionable.\par
\begin{figure}[ht]
\centering
\boxed{\includegraphics[width=0.97\textwidth]{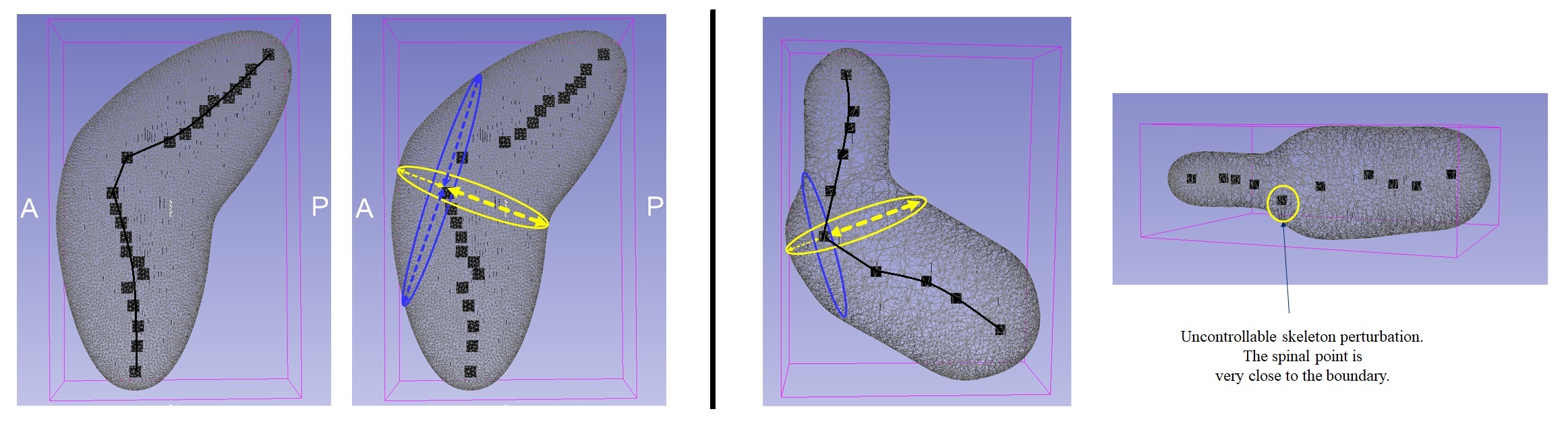}}
\caption{Visualisation of the spine of two 3EOs based on the local separator skeleton. The black curve represents the smooth version of the spine. The spines of both objects are bent and swung toward the left side of the objects.}
\label{fig:localSeparatorIssue}
\end{figure}

For a better intuition, we provide a very simple example to discuss the issue based on skeletonization via local separators of \citet{baerentzen2021skeletonization, baerentzen2023multilevel}. In simple words, the method searches for closed rings called local separators on the boundary. The center points of local separators define the skeleton. \Cref{fig:localSeparatorIssue} shows the spine of two 3EOs with smooth boundaries. The obtained spines are based on the PyGEL library (\url{https://www2.compute.dtu.dk/projects/GEL/PyGEL}). The spine of each object is not satisfactory as it is non-smooth and perturbed. By considering a smooth version of the spine as depicted by a black curve, the spine still suffers from critical issues. It seems the spine swings and bends freely towards the left parts of the objects. Probably the reason is that the algorithm detects the local separator as the blue curve rather than the yellow curve. On the other hand, the spine violates the RCC.

\subsection*{CVS relationship with medial skeleton}
\begin{figure}[ht]
\centering
\boxed{\includegraphics[width=0.60\textwidth]{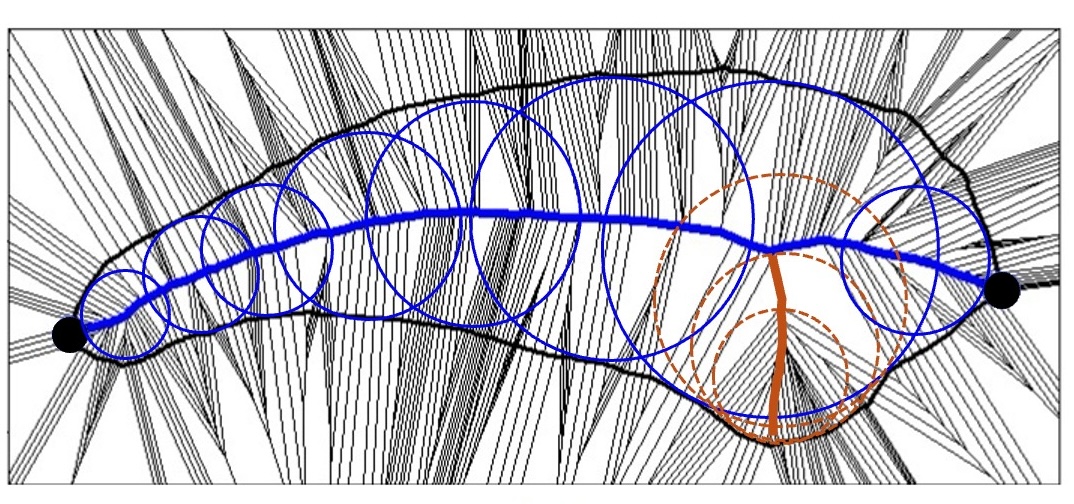}}
\caption{Approximation of the medial axis (in blue and red) using a Voronoi diagram. The two vertices (black dots) divide the boundary into two upper and lower components. The CVS is shown as the blue curve connecting the two vertices, representing the centers of the (blue) circles tangent to both the upper and lower components. The red curve represents the centers of the (dashed red) circles tangent only to the lower boundary component.}
\label{fig:CSV_vs_Medial}
\end{figure}
The Central Voronoi Skeleton (CVS) can be viewed as an approximation of a portion of the medial skeleton of an ellipsoidal as the set of centers of maximal inscribed spheres that are simultaneously tangent to both the upper and lower boundary components, as illustrated in \Cref{fig:CSV_vs_Medial}.

\subsection*{Irregular 3EO}
\begin{figure}[ht]
\centering
\boxed{\includegraphics[width=0.97\textwidth]{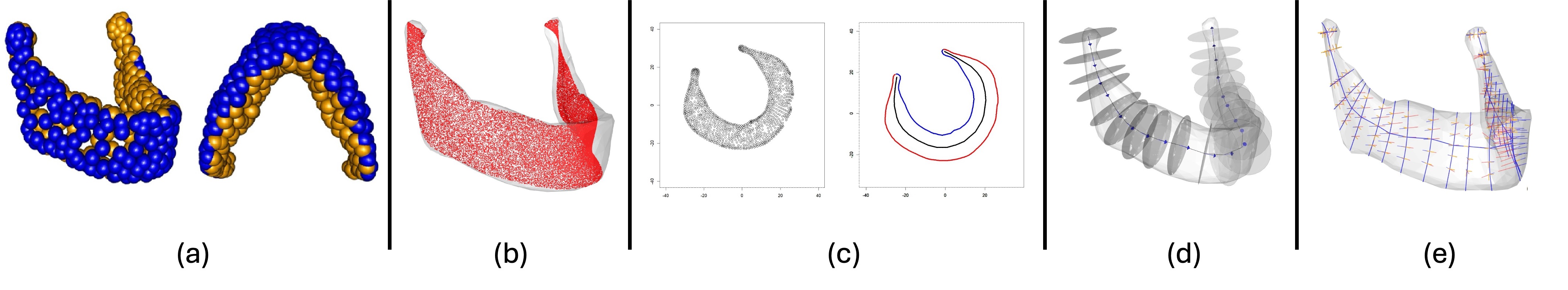}}
\caption{(a) Mandible top and bottom parts in blue and yellow, respectively. (b) Skeletal sheet of a mandible (without coronoid processes). (c) Flattened skeletal sheet by t-SNE (left), and center curve of the flattened skeletal sheet (right). (d) Slicing planes of the mandible along the spine. (e) Fitted LPDSSRep.}
\label{fig:mandibleCenterCurve}
\end{figure}
Theoretically, the skeletal sheet of an irregular 3EO can be bent and twisted with an extremely curvy structure. However, there are irregular 3EOs, such as the mandible, whose skeletal sheet has low local curvature. Thus, the skeletal sheet can be properly flattened. We consider such irregular 3EOs as \textit{treatable 3EOs} if the flattened skeletal sheet can be seen as a 2EO. Therefore, analogous to a regular 3EO, we can consider the inverse map of the center curve of the flattened skeletal sheet, i.e., $\mathscr{F}^{-1}(\Gamma'_2)$ of a treatable 3EO as the spine of the object, where $\mathscr{F}$ is a proper embedding. In this case, we define cross-sections normal to the spine.\par
From initial studies, we found the \textit{t-distributed stochastic neighbor embedding} (t-SNE) method \citep{van2014accelerating} suitable for flattening the skeletal sheet of treatable 3EOs. The t-SNE maps high-dimensional data to a lower-dimensional space in such a way that preserves the local structure and relationships between data points as much as possible. It is similar to the SNE of \citet{hinton2002stochastic}, but instead of using a Gaussian distribution to model these relationships, it employs the t-distribution \citep{bunte2012stochastic}. \Cref{fig:mandibleCenterCurve} illustrates the top and bottom parts of a mandible (a), its skeletal sheet (b), its t-SNE flattened version as a 2EO plus the center curve in 2D (c), the mandible with the normal slicing planes along its spine (d), and the fitted LPDSSRep (e).

Often, the complexity of the model fitting problem can be reduced by partitioning an irregular 3EO into several regular 3EOs. For example, skeletal analysis of a mandible can be based on skeletal models of two segmented hemimandibles \citep{alhadidi20123d} as regular 3EOs (see \Cref{fig:caudateHippo3DSpokes} (right)).

\subsection*{2EO straightening}
\begin{figure}[ht]
\centering
\boxed{\includegraphics[width=0.70\textwidth]{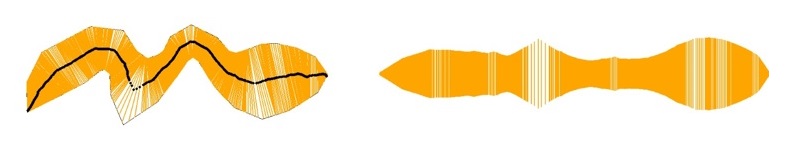}}
\caption{Illustration of 2EO straightening based on its chordal structure.}
\label{fig:straightening2DTube}
\end{figure}

As depicted in \Cref{fig:straightening2DTube}, we can use the chordal structure of a 2EO to straighten an object. Let $\{c_i\}_{i=1}^n$ be $n$ consecutive chords (i.e., cross-sections) with points $\{\bm{p}_i\}_{i=1}^n$, such that $\forall{i}$; $\bm{p}_i$ is the middle of $c_i$. Let $\|c_i\|$ be the length of the $i$th chord. To accomplish the object straightening we use a sequence of points $\{\bm{p}'_i\}_{i=1}^n$ corresponding to $\{\bm{p}_i\}_{i=1}^n$ on a straight line $\ell$ such that $\forall{i,j}$, $\|\bm{p}'_i-\bm{p}'_j\|=d_g(\bm{p}_i,\bm{p}_j)$, and line segments $\{c'_i\}_{i=1}^n$ corresponding to $\{c_i\}_{i=1}^n$ such that $\forall{i}$, $c'_i$ is perpendicular to $\ell$, the middle of $c'_i$ is on $\ell$, and $\|c'_i\|=\|c_i\|$. 

\subsection*{Classification}
For classification, we follow the idea of \textit{Composite PNS} (CPNS) \citet{pizer2013nested}. We normalize and vectorize the Euclideanized LPDSSReps (or LPDSReps). Thus, the feature space becomes $\mathbb{R}^{2n_c+9n_f}$ (or  $\mathbb{R}^{2n_c+9n_f+1}$ in case we consider scaled LPDSSReps and include the LP-size). In this sense, the problem becomes a binary classification problem of two groups of vectors that can be solved by, e.g., \textit{Support Vector Machine} (SVM)\citep{noble2006support}.\par
For the classification of the real data from \Cref{sec:Statistical_analysis}, we applied different standard binary classification methods including \textit{K Nearest Neighbor} (KNN), SVM with linear and radial kernels, and \textit{Naive Bayesian} (NB) classification. Further, we applied 10-fold cross-validation to evaluate the accuracy, specificity, and sensitivity of the outcomes based on Cohen's kappa \citep{cohen1960coefficient} as shown in \Cref{Table_classification}. As we expected, the classification based on both LPDSRep and LPDSSRep is not promising because, basically, we do not expect to observe separable local distributions at the early stages of PD. However, overall, it seems that the LPDSSRep has a better performance compared to the LPDSRep according to both accuracy and Cohen's kappa.

\begin{table}[ht]
\caption{LPDSRep classification based on 10-fold cross-validation.}
\label{Table_classification}
\centering
\resizebox{\textwidth}{!}{%
\begin{tabular}{@{}ccccccc@{}}
\toprule
\multirow{3}{*}{\begin{tabular}[c]{@{}c@{}}Classes\end{tabular}} & \multirow{3}{*}{\begin{tabular}[c]{@{}c@{}}Model fitting\end{tabular}} & \multirow{3}{*}{\begin{tabular}[c]{@{}c@{}}Assessment\end{tabular}} & \multicolumn{4}{c}{Classification methods} \\ \cmidrule(l){4-7} 
 &  &  & KNN & \begin{tabular}[c]{@{}c@{}}SVM-linear \end{tabular} & \begin{tabular}[c]{@{}c@{}}SVM-radial \end{tabular} & NB \\ \midrule
\multirow{6}{*}{\begin{tabular}[c]{@{}c@{}} PD \& CG\end{tabular}} & \multirow{2}{*}{LPDSRep} & Accuracy & 0.60 & 0.55 & 0.63 & 0.57 \\ \cmidrule(l){3-7} 
 &  & Kappa & 0.10 & 0.05 & 0.06 & 0.14 \\ \cmidrule(l){2-7} 
 & \multirow{2}{*}{LPDSSRep} & Accuracy & 0.63 & 0.63 & 0.60 & 0.61 \\
 \cmidrule(l){3-7} 
 &  & Kappa & 0.18 & 0.17 & 0.10 & 0.11 \\
 \bottomrule
\end{tabular}%
}
\end{table}


\section*{Declarations}
\subsection*{Ethical Approval}
Not applicable.
\subsection*{Competing interests}
The authors declare no competing interests.
\subsection*{Authors' contributions}
Mohsen Taheri is the first author and writer of the manuscript. Jörn Schulz and Stephen Pizer have supervised the project. All authors read and approved the final manuscript.
\subsection*{Funding}
Not applicable
\subsection*{Availability of data and materials}
Implementation of the manuscript's methodology is available as R scripts on \url{https://github.com/MohsenTaheriShalmani/LP-dss-rep}.
Due to limited permission to share the Parkwest data, we only provide R code for producing toy examples and synthetic data in the repository.

\bibliographystyle{spbasic}
\bibliography{bibliography}

\end{document}